\documentclass[letterpaper,twocolumn,10pt]{article}\usepackage[]{graphicx}\usepackage[]{color}
\makeatletter
\def\maxwidth{ %
  \ifdim\Gin@nat@width>\linewidth
    \linewidth
  \else
    \Gin@nat@width
  \fi
}
\makeatother

\definecolor{fgcolor}{rgb}{0.345, 0.345, 0.345}

\usepackage{framed}
\makeatletter
 {\par\unskip\endMakeFramed%
 \at@end@of@kframe}
\makeatother

\definecolor{shadecolor}{rgb}{.97, .97, .97}
\definecolor{messagecolor}{rgb}{0, 0, 0}
\definecolor{warningcolor}{rgb}{1, 0, 1}
\definecolor{errorcolor}{rgb}{1, 0, 0}

\usepackage{alltt}
\usepackage{usenix}

\usepackage[letterpaper]{geometry}
\usepackage{epsfig,endnotes,url}

\usepackage{xspace}

\usepackage[HideComments]{mycomment}

\usepackage[english]{babel}
\usepackage[latin1]{inputenc}
\usepackage{amsmath,amsthm, amssymb, latexsym}

\usepackage{graphicx}
\usepackage{multirow}
\usepackage{booktabs}

\usepackage{caption,subcaption}

\usepackage{paralist}

\usepackage{balance}

\usepackage{afterpage}

\usepackage{amssymb}
\usepackage{tabularx,colortbl}
\usepackage{eurosym}
\usepackage{xcolor}
\usepackage{pgfplotstable}
\usepackage{tikz}
\usepackage{pgfplots}
\usepackage{pgf-umlsd}
\usetikzlibrary{shapes,arrows}
\definecolor{findOptimalPartition}{HTML}{696969}
\definecolor{storeClusterComponent}{HTML}{808080}
\definecolor{dbscan}{HTML}{BEBEBE}
\definecolor{constructCluster}{HTML}{DCDCDC}

\usepackage{versions}
\excludeversion{DocumentVersionSTAST}
\includeversion{DocumentVersionTR}

\newtheorem{researchquestion}{RQ}

\newcommand{\const}[1]{\ensuremath{\mathsf{#1}\xspace}}

\begin{DocumentVersionSTAST}
\copyrightyear{2018} 
\acmYear{2018} 
\setcopyright{acmcopyright}
\acmConference[STAST2018]{8th International Workshop on Socio-Technical Aspects in Security and Trust}{December 4, 2018}{San Juan, Puerto Rico, USA}


\end{DocumentVersionSTAST}

\IfFileExists{upquote.sty}{\usepackage{upquote}}{}
\begin{document}




\newcommand{\regTableChallenged}{
\begin{table*}[ht]
\centering
\caption{Logistic Regression: User Challenged} 
\label{tab:regTableChallenged}
\begingroup\footnotesize
\begin{tabular}{rrrrr}
  \hline
 & Estimate & Std. Error & z value & Pr($>$$|$z$|$) \\ 
  \hline
(Intercept) & -2.981 & 1.021 & -2.921 & 0.003 \\ 
  Machine.Data & 1.893 & 0.727 & 2.602 & 0.009 \\ 
  Value & 1.498 & 0.705 & 2.125 & 0.034 \\ 
  Region & 1.893 & 0.727 & 2.602 & 0.009 \\ 
  Website & -0.219 & 0.663 & -0.330 & 0.741 \\ 
  Card & -0.563 & 0.312 & -1.805 & 0.071 \\ 
   \hline
\end{tabular}
\endgroup
\end{table*}
}

\newcommand{\tabRegTableChallengedRefined}{
\begin{table*}[tbp]
\centering
\caption{Logistic Regression: User Challenged} 
\label{tab:regTableChallengedRefined}
\begin{tabular}{rrrrl}
  \toprule
 & Estimate & SE & $z$-value & $p$-Value \\ 
  \midrule
(Intercept) & -2.981 & 1.021 & -2.921 & $\phantom{<}.003**$ \\ 
  Machine.Data & 1.893 & 0.727 & 2.602 & $\phantom{<}.009**$ \\ 
  Value & 1.498 & 0.705 & 2.125 & $\phantom{<}.034*$ \\ 
  Region & 1.893 & 0.727 & 2.602 & $\phantom{<}.009**$ \\ 
  Website & -0.219 & 0.663 & -0.330 & $\phantom{<}.741$ \\ 
  Card & -0.563 & 0.312 & -1.805 & $\phantom{<}.071$ \\ 
   \bottomrule
 \multicolumn{5}{c}{\emph{Note:} Overall Model: Wald $\chi^2(5) = 21.593$, $p < .001$}\\
 \multicolumn{5}{c}{$R^2$= .28 (Hosmer \& Lemeshow), .29 (Cox \& Snell), .41 (Nagelkerke)}\\
\end{tabular}
\end{table*}
}

\newcommand{\regTableDeclined}{
\begin{table*}[ht]
\centering
\caption{Logistic Regression: Transaction Declined} 
\label{tab:regTableDeclined}
\begingroup\footnotesize
\begin{tabular}{rrrrr}
  \hline
 & Estimate & Std. Error & z value & Pr($>$$|$z$|$) \\ 
  \hline
(Intercept) & -6.333 & 1.709 & -3.706 & 0.000 \\ 
  Machine.Data & 2.944 & 0.944 & 3.119 & 0.002 \\ 
  Value & 3.397 & 0.996 & 3.410 & 0.001 \\ 
  Region & 3.397 & 0.996 & 3.410 & 0.001 \\ 
  Website & 0.285 & 0.757 & 0.376 & 0.707 \\ 
  Card & 0.975 & 0.398 & 2.449 & 0.014 \\ 
   \hline
\end{tabular}
\endgroup
\end{table*}
}

\newcommand{\tabRegTableDeclinedRefined}{
\begin{table*}[tbp]
\centering
\caption{Logistic Regression: Transaction Declined} 
\label{tab:regTableDeclinedRefined}
\begin{tabular}{rrrrl}
  \toprule
 & Estimate & SE & $z$-value & $p$-Value \\ 
  \midrule
(Intercept) & -6.333 & 1.709 & -3.706 & $<.001***$ \\ 
  Machine.Data & 2.944 & 0.944 & 3.119 & $\phantom{<}.002**$ \\ 
  Value & 3.397 & 0.996 & 3.410 & $<.001***$ \\ 
  Region & 3.397 & 0.996 & 3.410 & $<.001***$ \\ 
  Website & 0.285 & 0.757 & 0.376 & $\phantom{<}.707$ \\ 
  Card & 0.975 & 0.398 & 2.449 & $\phantom{<}.014*$ \\ 
   \bottomrule
 \multicolumn{5}{c}{\emph{Note:} Overall Model: Wald $\chi^2(5) = 44.409$, $p < .001$}\\
 \multicolumn{5}{c}{$R^2$= .50 (Hosmer \& Lemeshow), .50 (Cox \& Snell), .67 (Nagelkerke)}\\
\end{tabular}
\end{table*}
}

\newcommand{\regTableBlocked}{
\begin{table*}[ht]
\centering
\caption{Logistic Regression: Card Blocked} 
\label{tab:regTableBlocked}
\begingroup\footnotesize
\begin{tabular}{rrrrr}
  \hline
 & Estimate & Std. Error & z value & Pr($>$$|$z$|$) \\ 
  \hline
(Intercept) & -22.621 & 2813.652 & -0.008 & 0.994 \\ 
  Machine.Data & 1.144 & 0.911 & 1.255 & 0.209 \\ 
  Value & 20.291 & 2813.651 & 0.007 & 0.994 \\ 
  Region & 2.428 & 0.986 & 2.462 & 0.014 \\ 
  Website & 0.388 & 0.886 & 0.438 & 0.661 \\ 
  Card & -0.386 & 0.404 & -0.955 & 0.339 \\ 
   \hline
\end{tabular}
\endgroup
\end{table*}
}

\newcommand{\tabRegTableBlockedRefined}{
\begin{table*}[tbp]
\centering
\caption{Logistic Regression: Card Blocked} 
\label{tab:regTableBlockedRefined}
\begin{tabular}{rrrrl}
  \toprule
 & Estimate & SE & $z$-value & $p$-Value \\ 
  \midrule
(Intercept) & -22.621 & 2813.652 & -0.008 & $\phantom{<}.994$ \\ 
  Machine.Data & 1.144 & 0.911 & 1.255 & $\phantom{<}.209$ \\ 
  Value & 20.291 & 2813.651 & 0.007 & $\phantom{<}.994$ \\ 
  Region & 2.428 & 0.986 & 2.462 & $\phantom{<}.014*$ \\ 
  Website & 0.388 & 0.886 & 0.438 & $\phantom{<}.661$ \\ 
  Card & -0.386 & 0.404 & -0.955 & $\phantom{<}.339$ \\ 
   \bottomrule
 \multicolumn{5}{c}{\emph{Note:} Overall Model: Wald $\chi^2(5) = 27.497$, $p < .001$}\\
 \multicolumn{5}{c}{$R^2$= .47 (Hosmer \& Lemeshow), .35 (Cox \& Snell), .58 (Nagelkerke)}\\
\end{tabular}
\end{table*}
}

\newcommand{\tabRegTableBlockedRefinedInt}{
\begin{table*}[tbp]
\centering
\caption{Logistic Regression: Card Blocked} 
\label{tab:regTableBlockedRefinedInt}
\begin{tabular}{rrrrl}
  \toprule
 & Estimate & SE & $z$-value & $p$-Value \\ 
  \midrule
(Intercept) & -22.798 & 2855.831 & -0.008 & $\phantom{<}.994$ \\ 
  Value & 20.194 & 2855.830 & 0.007 & $\phantom{<}.994$ \\ 
  Region & 2.327 & 0.951 & 2.447 & $\phantom{<}.014*$ \\ 
  Machine.Data & 1.096 & 0.888 & 1.234 & $\phantom{<}.217$ \\ 
   \bottomrule
 \multicolumn{5}{c}{\emph{Note:} Overall Model: Wald $\chi^2(3) = 26.358$, $p < .001$}\\
 \multicolumn{5}{c}{$R^2$= .45 (Hosmer \& Lemeshow), .34 (Cox \& Snell), .56 (Nagelkerke)}\\
\end{tabular}
\end{table*}
}

\newcommand{\figRegPredChallenged}{
\begin{figure*}[p]
{\centering
\begin{subfigure}[b]{0.32\textwidth}

\includegraphics[width=\maxwidth]{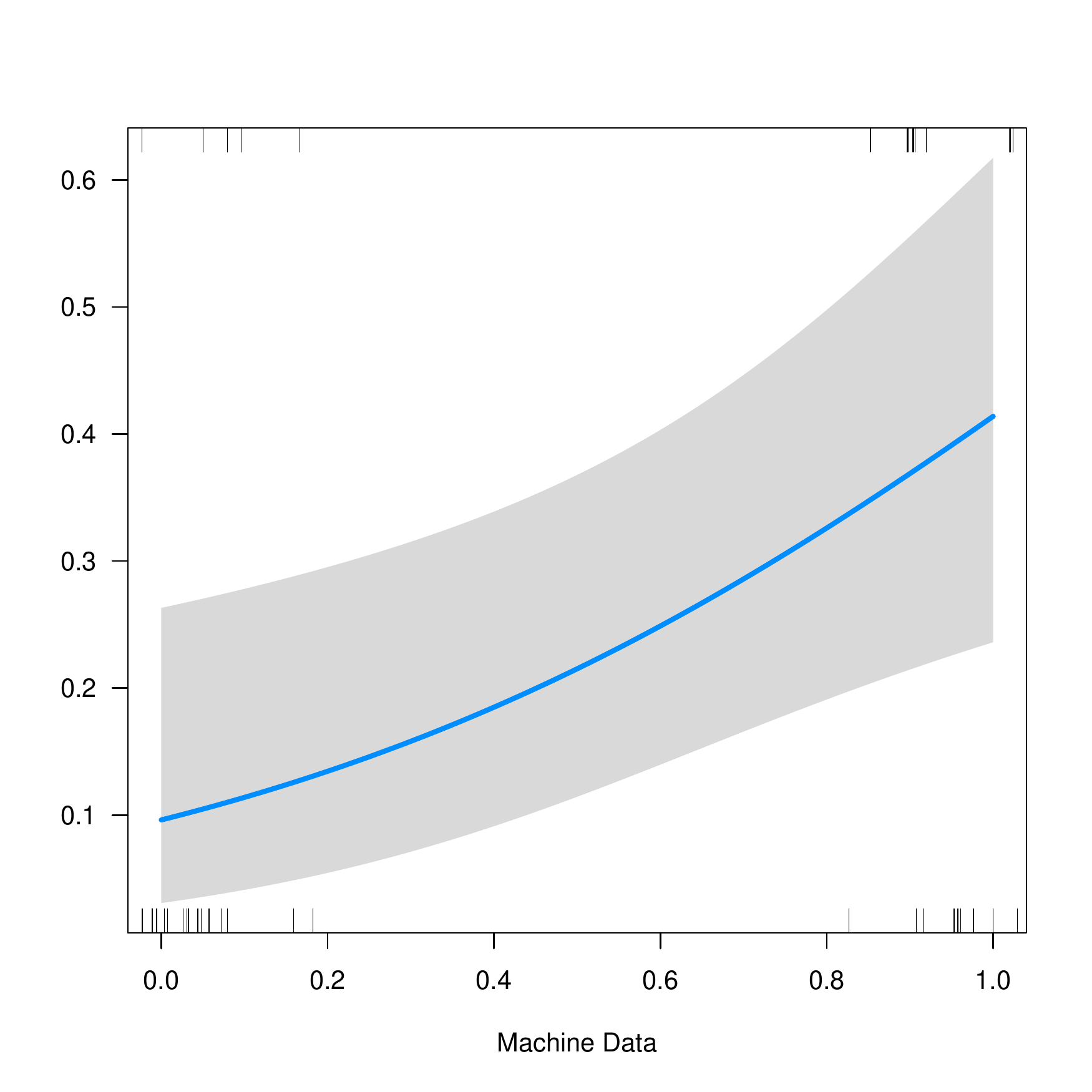} 
\subcaption{Change of Machine Data}
\end{subfigure}
\begin{subfigure}[b]{0.32\textwidth}

\includegraphics[width=\maxwidth]{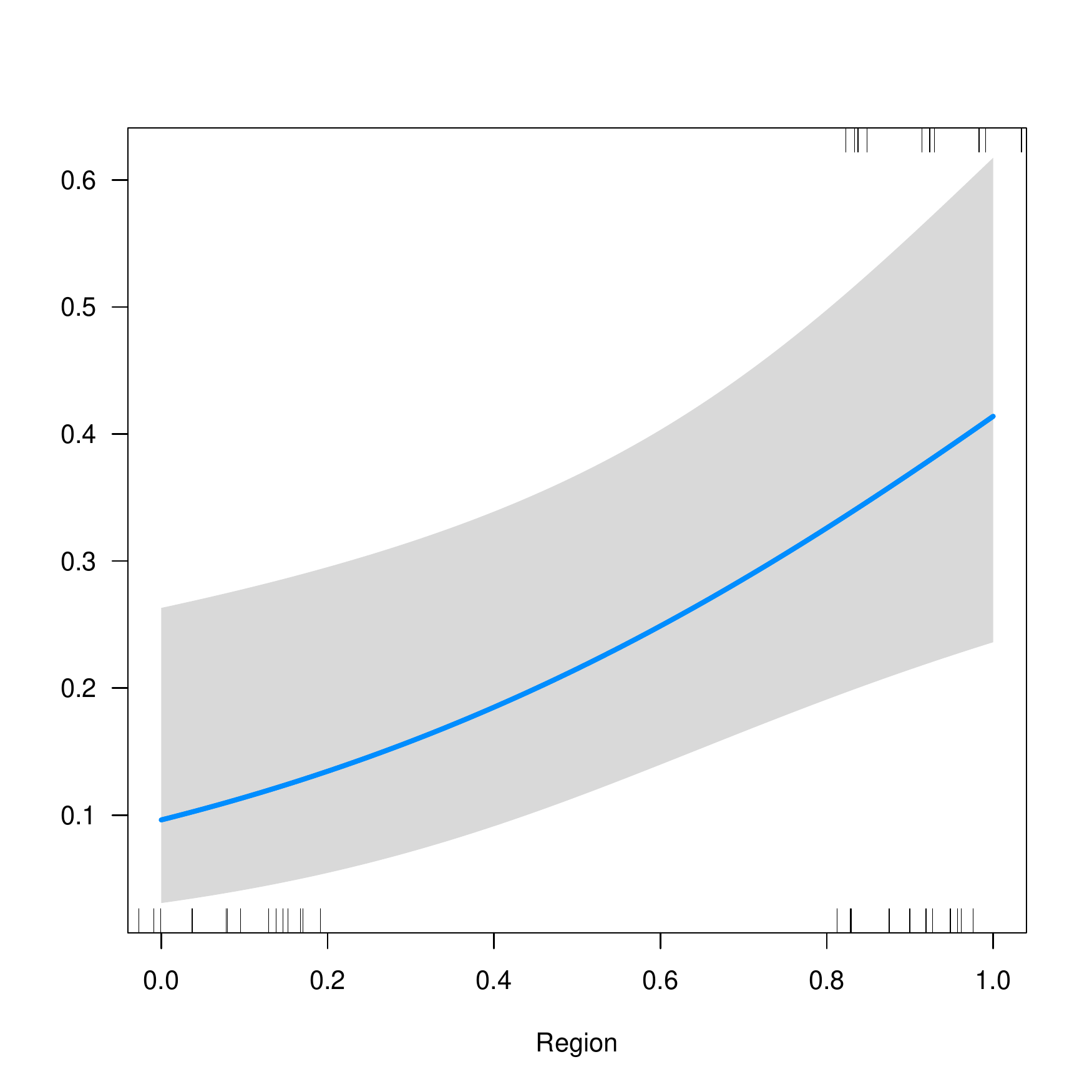} 
\subcaption{Change of Region}
\end{subfigure}
\begin{subfigure}[b]{0.32\textwidth}

\includegraphics[width=\maxwidth]{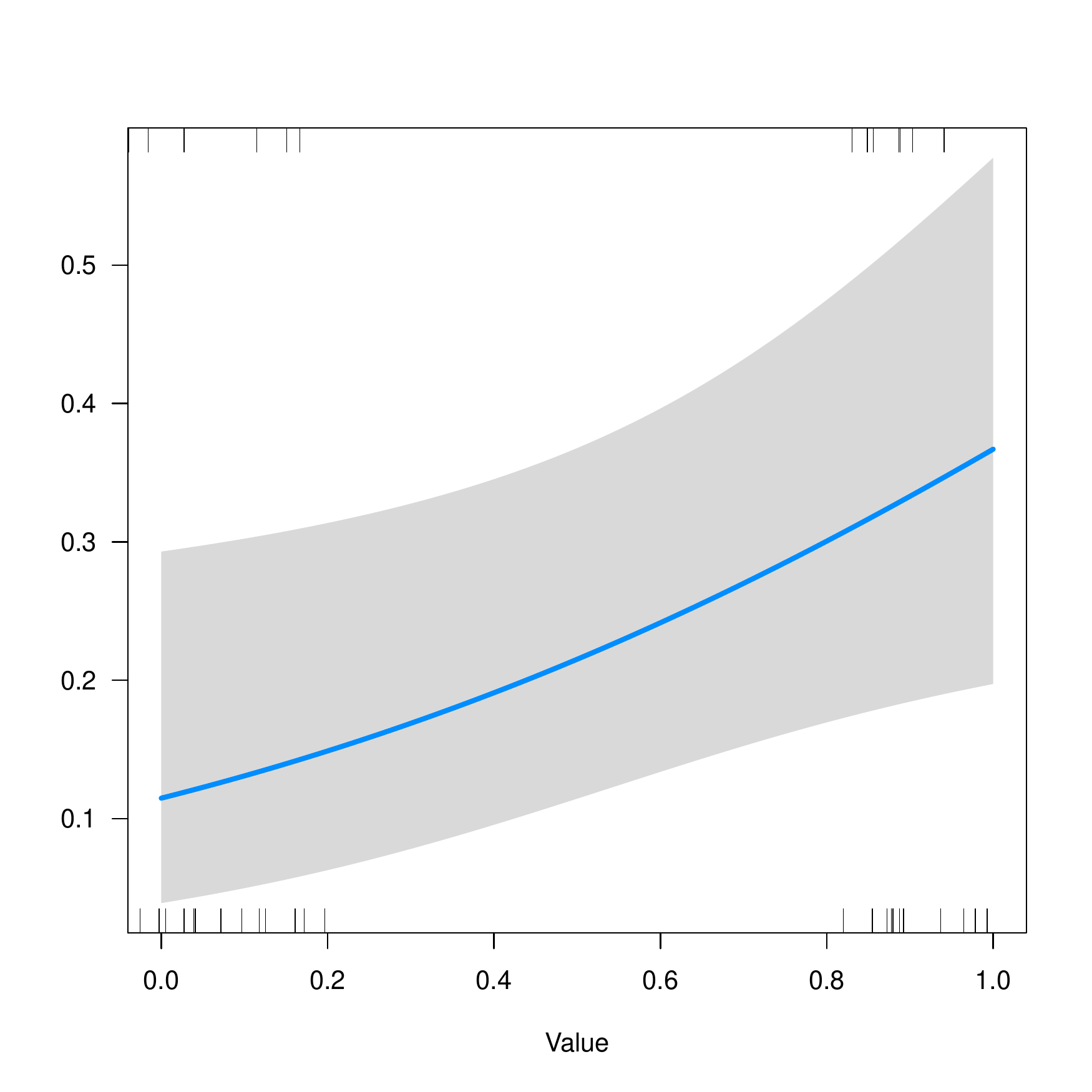} 
\subcaption{Change of Value}
\end{subfigure}
\caption{Probability(\textsf{challenged}) depending on significant predictors.}
\label{fig:RegPredChallenged}
}
\end{figure*}
}

\newcommand{\figRegPredDeclined}{
\begin{figure*}[p]
{\centering
\begin{subfigure}[b]{0.32\textwidth}

\includegraphics[width=\maxwidth]{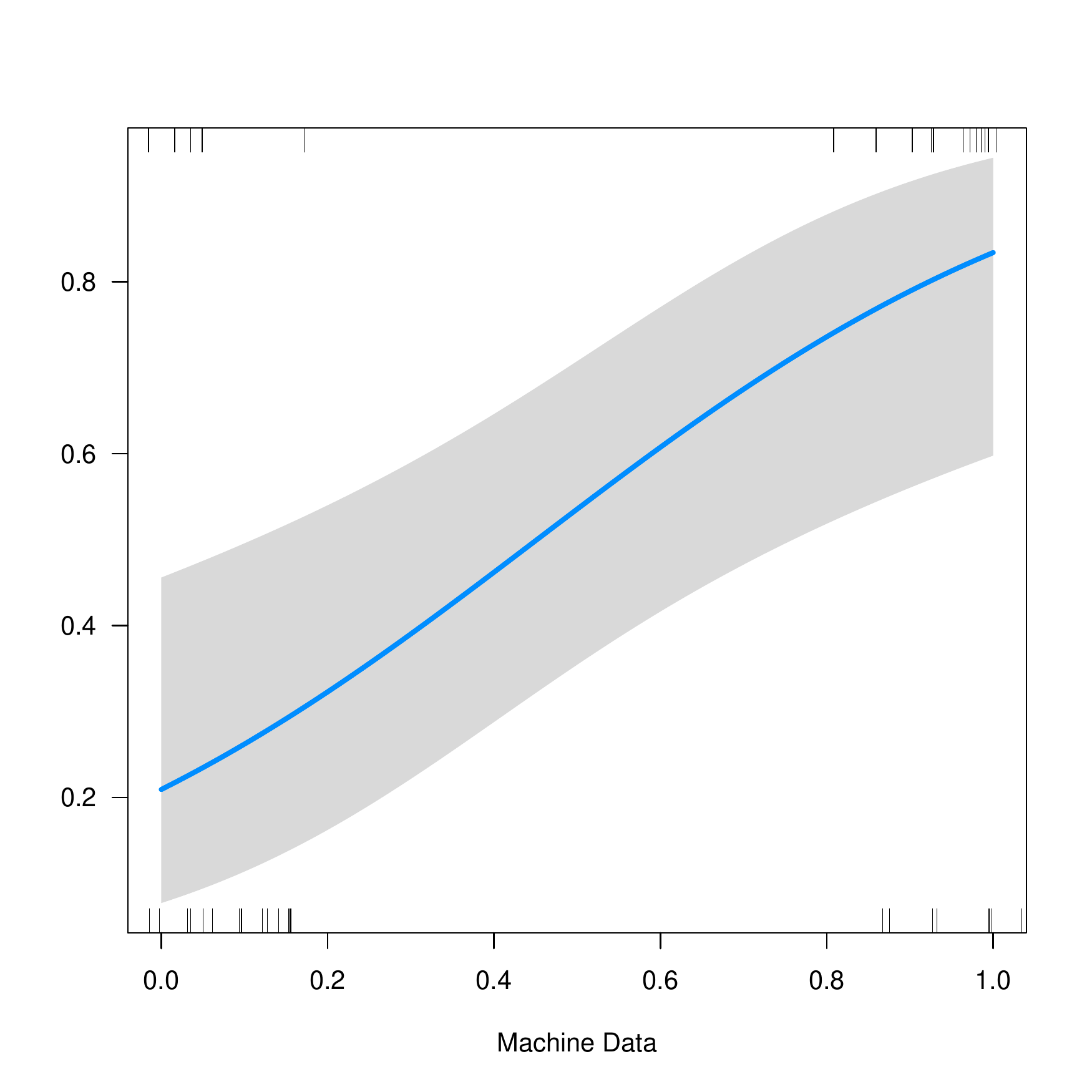} 
\subcaption{Change of Machine Data}
\end{subfigure}
\begin{subfigure}[b]{0.32\textwidth}

\includegraphics[width=\maxwidth]{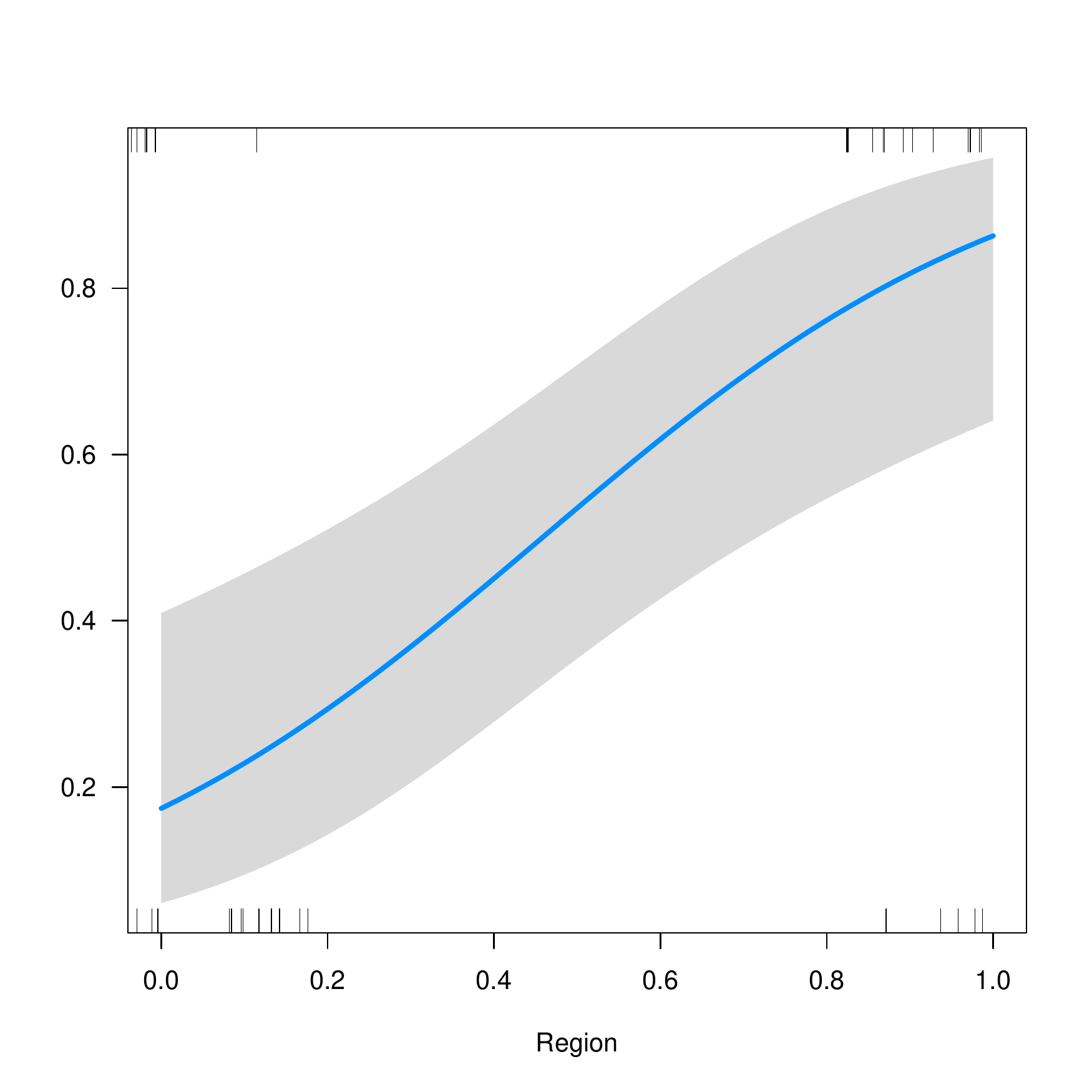} 
\subcaption{Change of Region}
\end{subfigure}
\begin{subfigure}[b]{0.32\textwidth}

\includegraphics[width=\maxwidth]{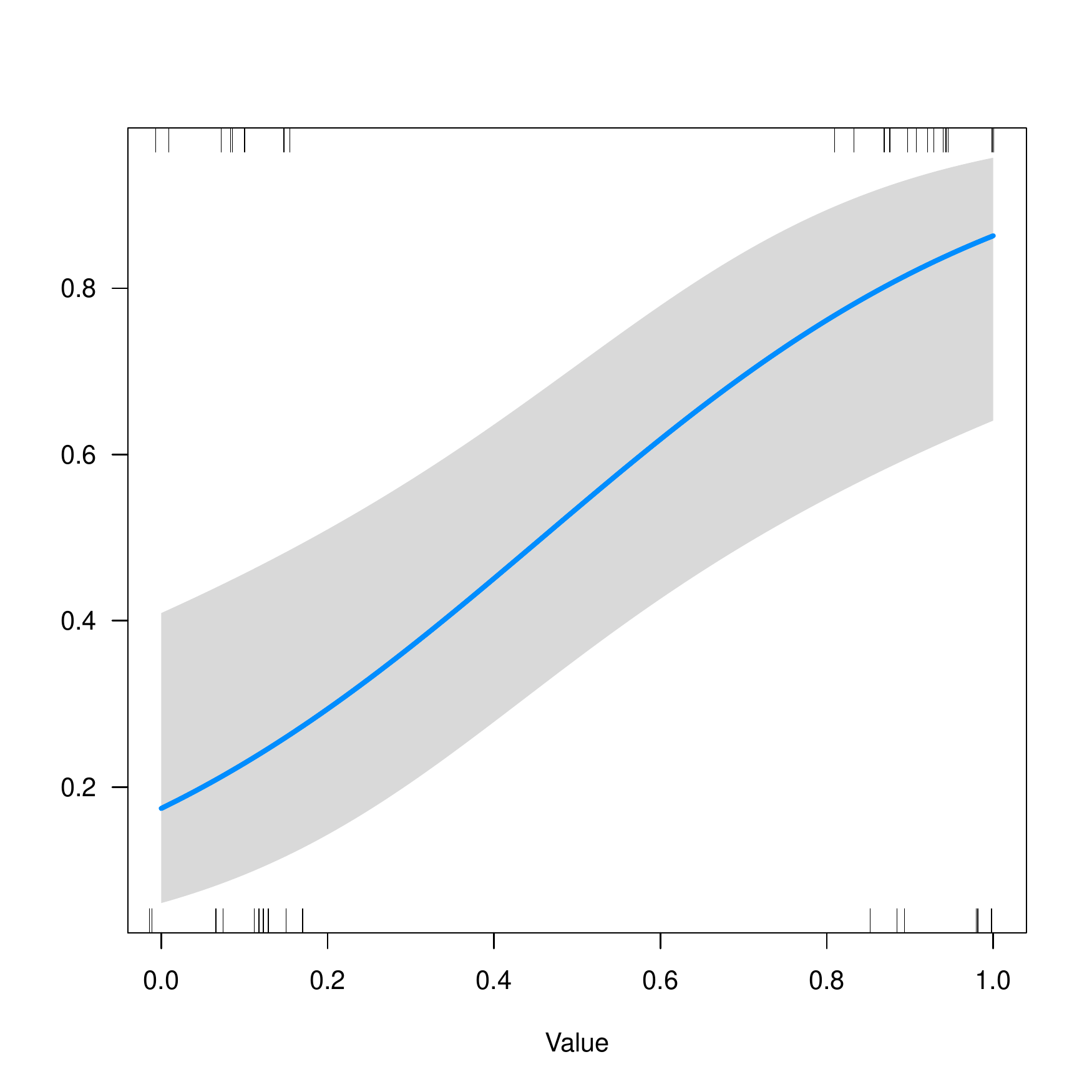} 
\subcaption{Change of Value}
\end{subfigure}
\caption{Probability(\textsf{declined}) depending on significant predictors.}
\label{fig:RegPredDeclined}
}
\end{figure*}
}

\newcommand{\figRegPredBlocked}{
\begin{figure}[tb]
{\centering

\includegraphics[width=\maxwidth]{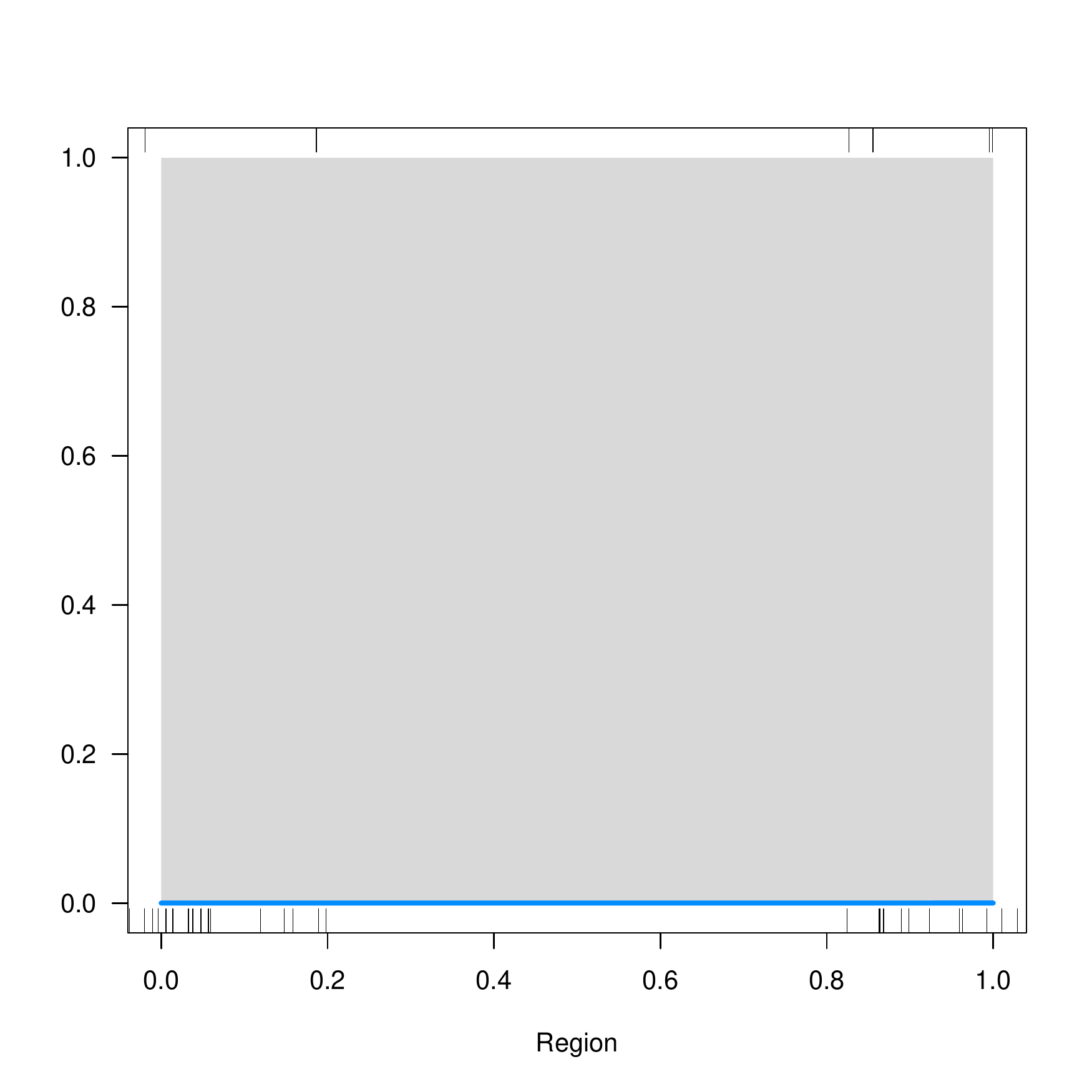} 
\caption{Probability(\textsf{blocked}) depending on the significant predictor \textsf{region}.}
\label{fig:RegPredBlocked}
}
\end{figure}
}

\newcommand{\figRegChallengedDataRegion}{
\begin{figure*}
{\centering
\begin{subfigure}[b]{0.49\textwidth}

\includegraphics[width=\maxwidth]{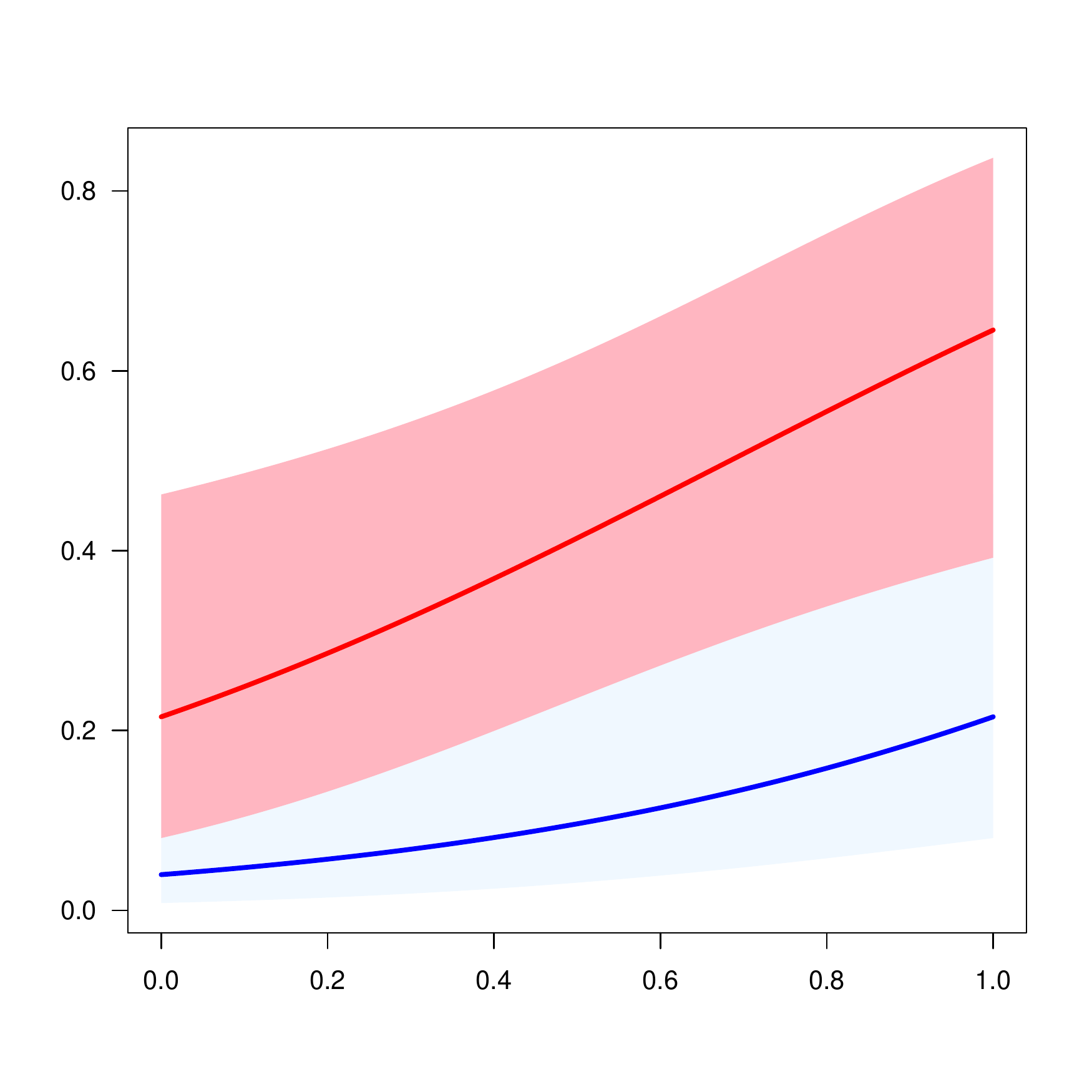} 
\subcaption{Overlay Plot \textsf{machinedata} by \textsf{region}}
\end{subfigure}
\begin{subfigure}[b]{0.49\textwidth}

\includegraphics[width=\maxwidth]{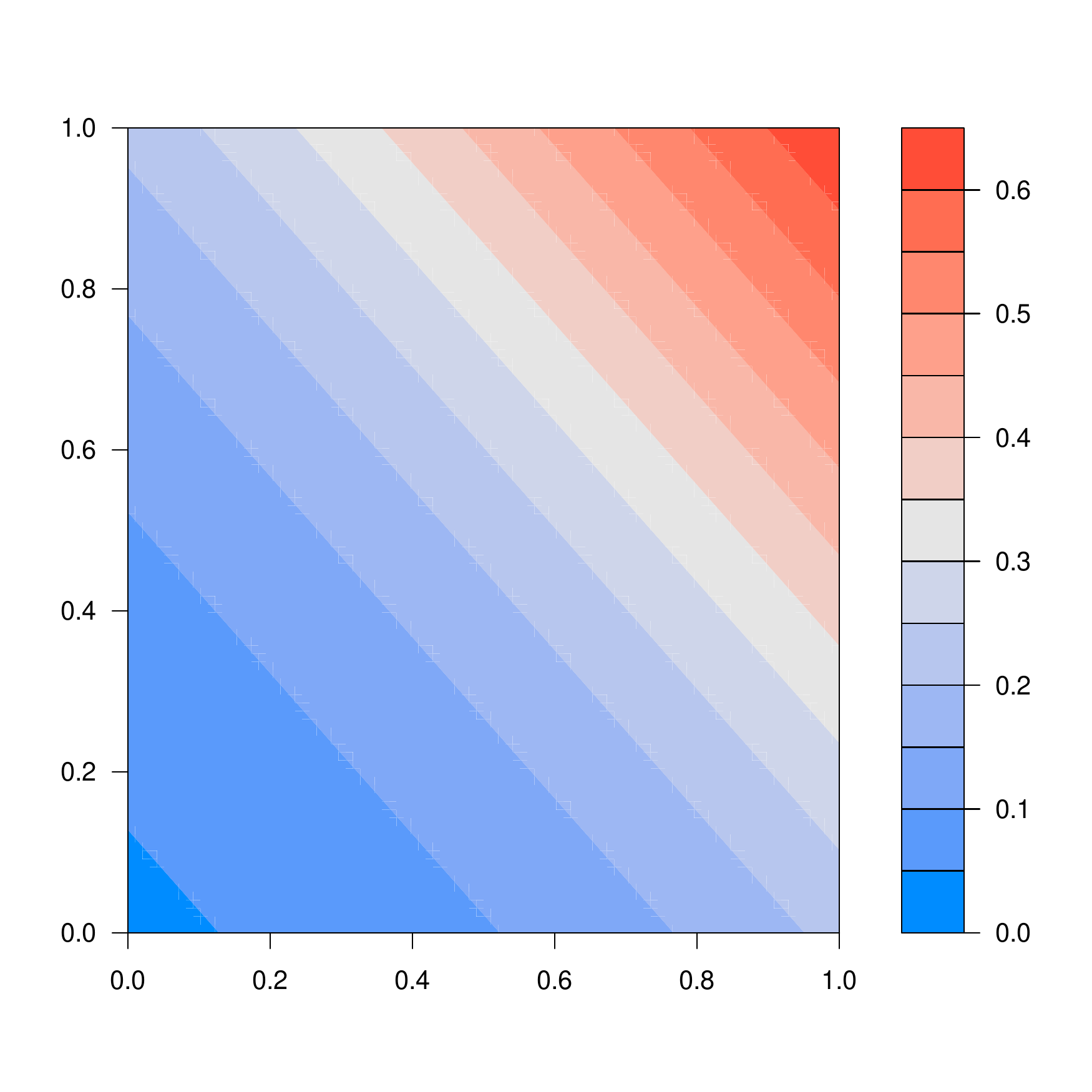} 
\subcaption{2-D Regression Surface}
\end{subfigure}
\caption{Probability(\textsf{challenged}) depending on machine data and region.}
\label{fig:regChallengedChallengedDataRegion}
}
\end{figure*}
}

\newcommand{\figRegChallengedValueRegion}{
\begin{figure*}
{\centering
\begin{subfigure}[b]{0.49\textwidth}

\includegraphics[width=\maxwidth]{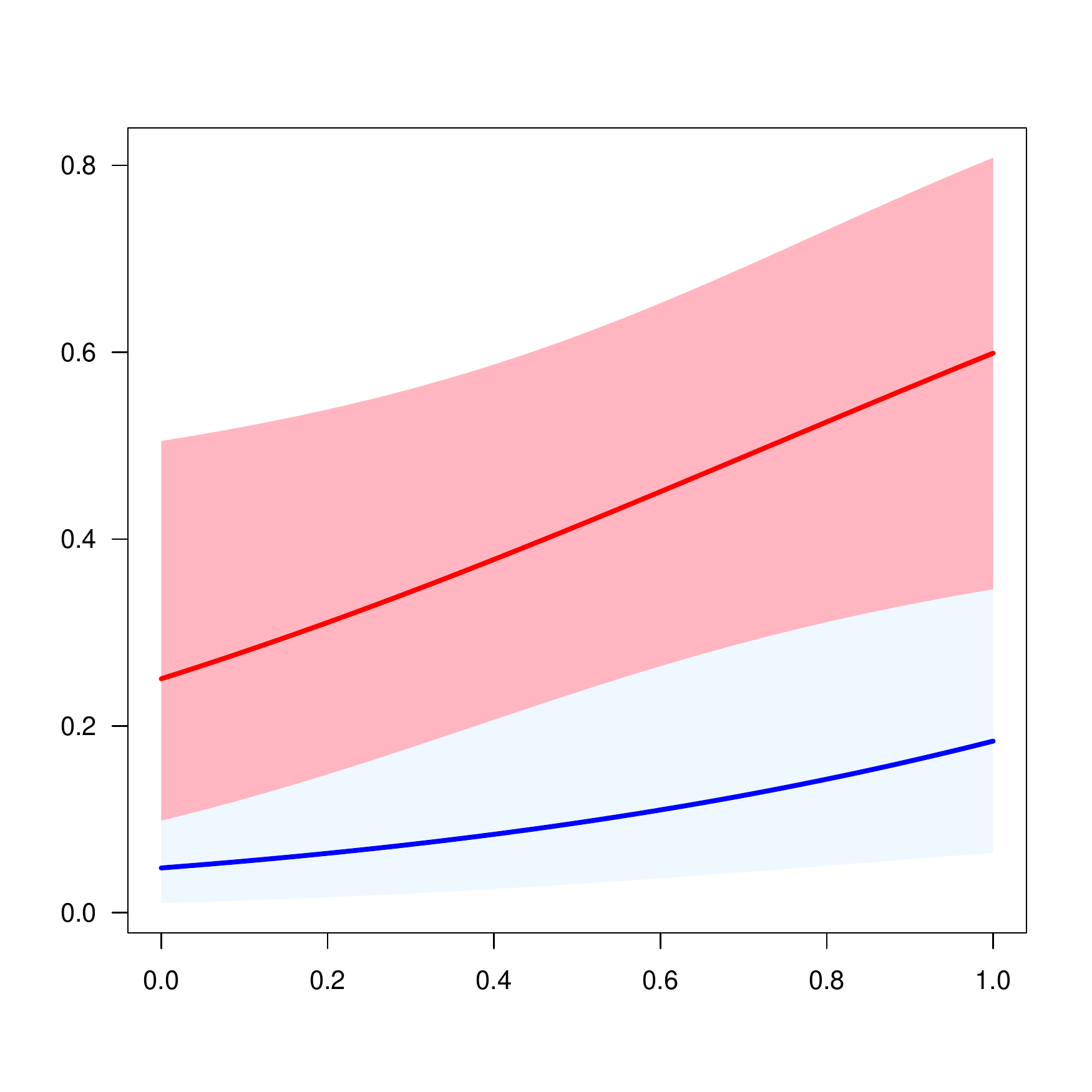} 
\subcaption{Overlay Plot \textsf{value} by \textsf{region}}
\end{subfigure}
\begin{subfigure}[b]{0.49\textwidth}

\includegraphics[width=\maxwidth]{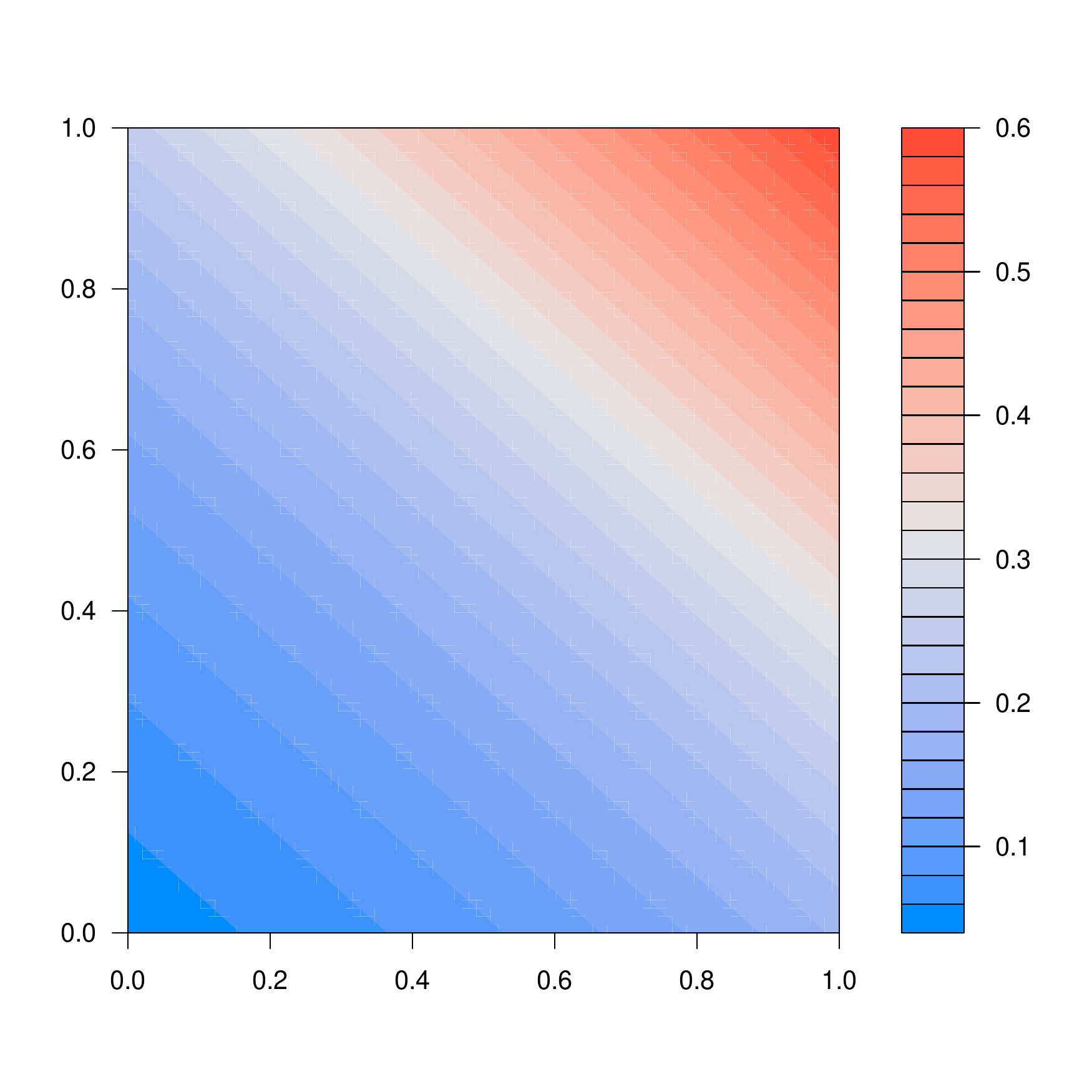} 
\subcaption{2-D Regression Surface}
\end{subfigure}
\caption{Probability(\textsf{challenged}) depending on transaction value and region.}
\label{fig:RegChallengedValueRegion}
}
\end{figure*}
}

\newcommand{\figRegChallengedDataValueRegion}{
\begin{figure*}[p]
{\centering
\begin{subfigure}[b]{0.49\textwidth}

\includegraphics[width=\maxwidth]{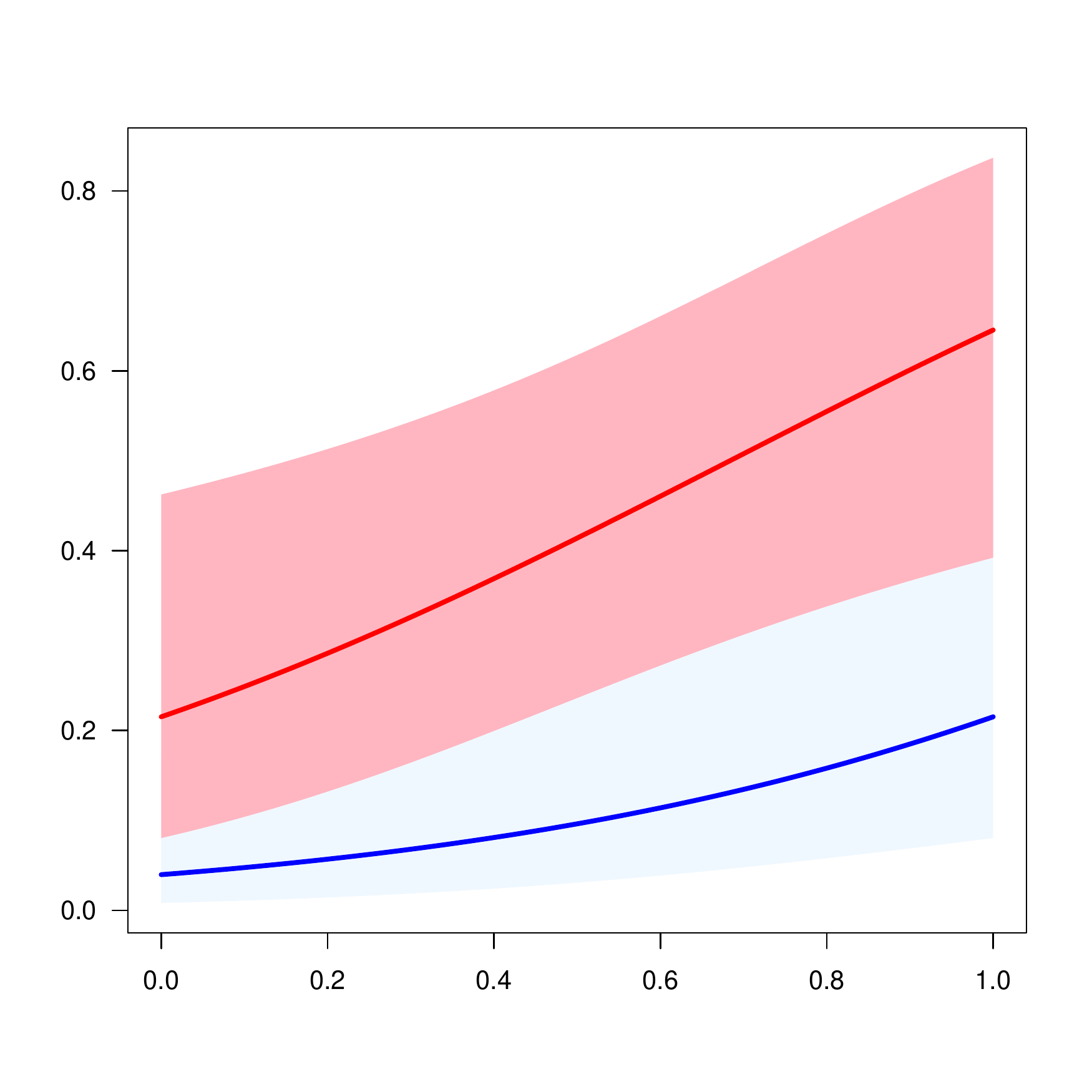} 
\vspace{-1cm}
\subcaption{\textsf{machine.data} by \textsf{region}}
\end{subfigure}
\begin{subfigure}[b]{0.49\textwidth}

\includegraphics[width=\maxwidth]{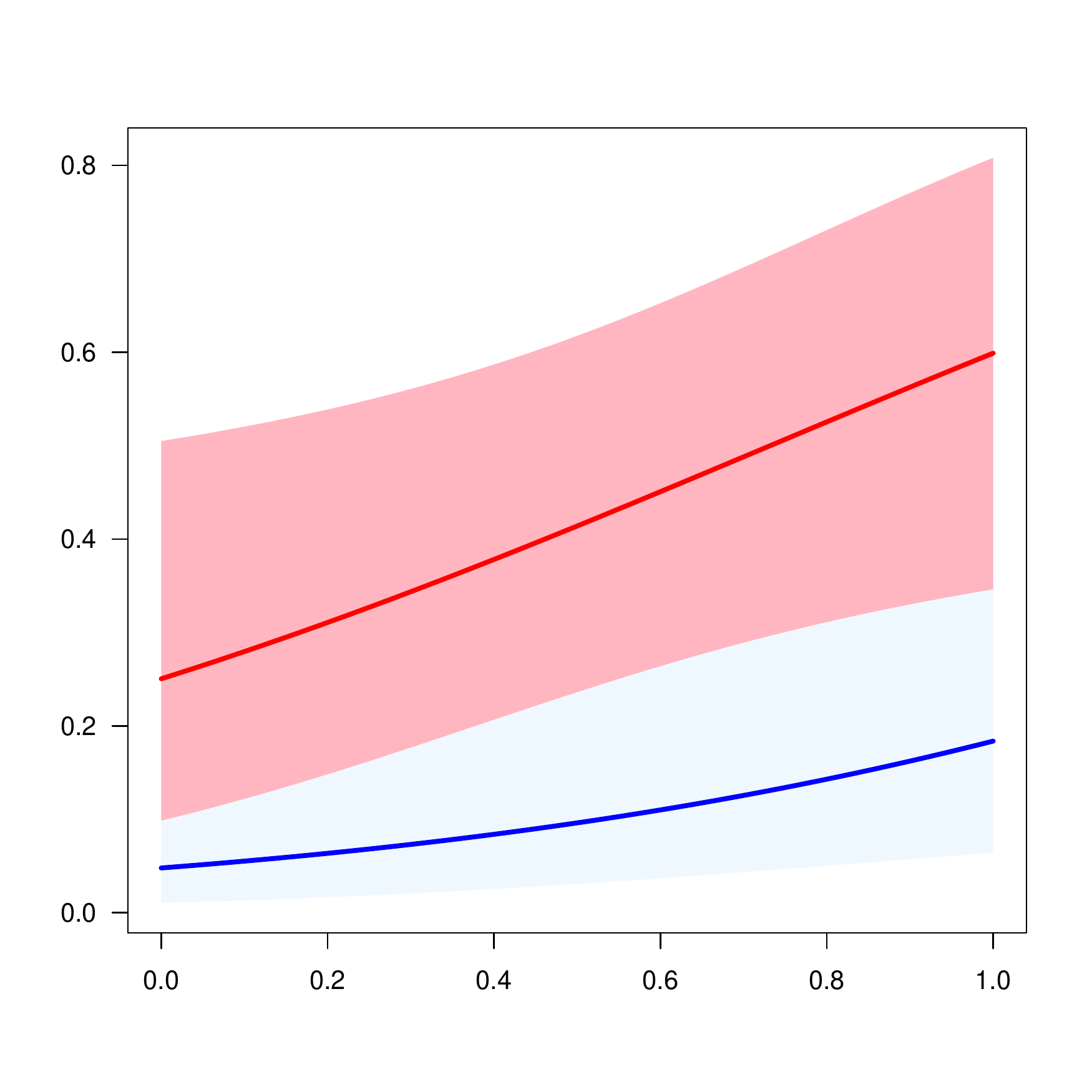} 
\vspace{-1cm}
\subcaption{\textsf{value} by \textsf{region}}
\end{subfigure}
\caption{Probability(\textsf{challenged}) depending on [machine data \emph{or} transaction value] by region.
           (The blue line on the bottom shows the home region, the red line on the top the foreign region)}
\label{fig:RegChallengedDataValueRegion}
}
\end{figure*}
}

\newcommand{\figegDeclinedDataRegion}{
\begin{figure*}
{\centering
\begin{subfigure}[b]{0.49\textwidth}

\includegraphics[width=\maxwidth]{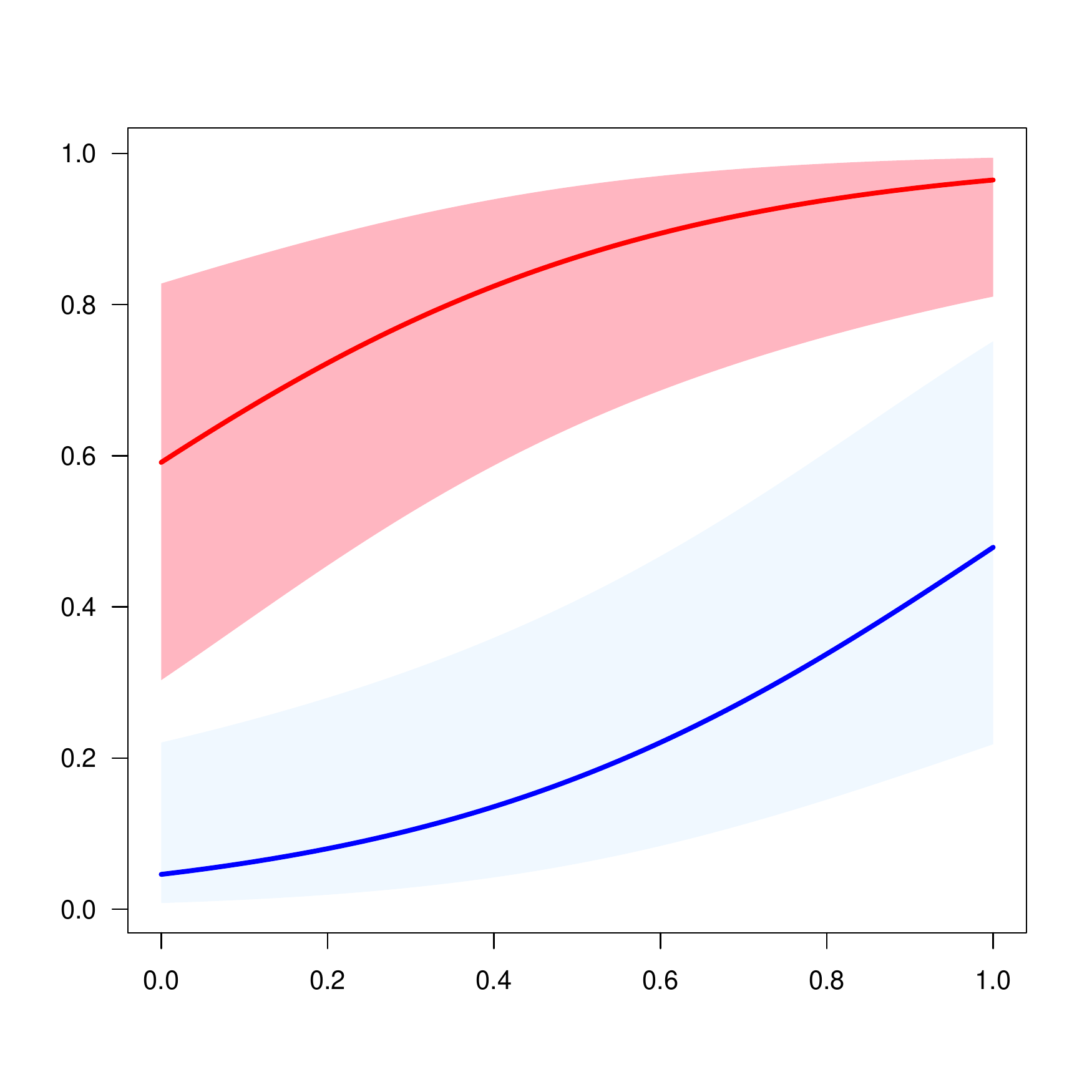} 
\subcaption{Overlay Plot \textsf{machinedata} by \textsf{region}}
\end{subfigure}
\begin{subfigure}[b]{0.49\textwidth}

\includegraphics[width=\maxwidth]{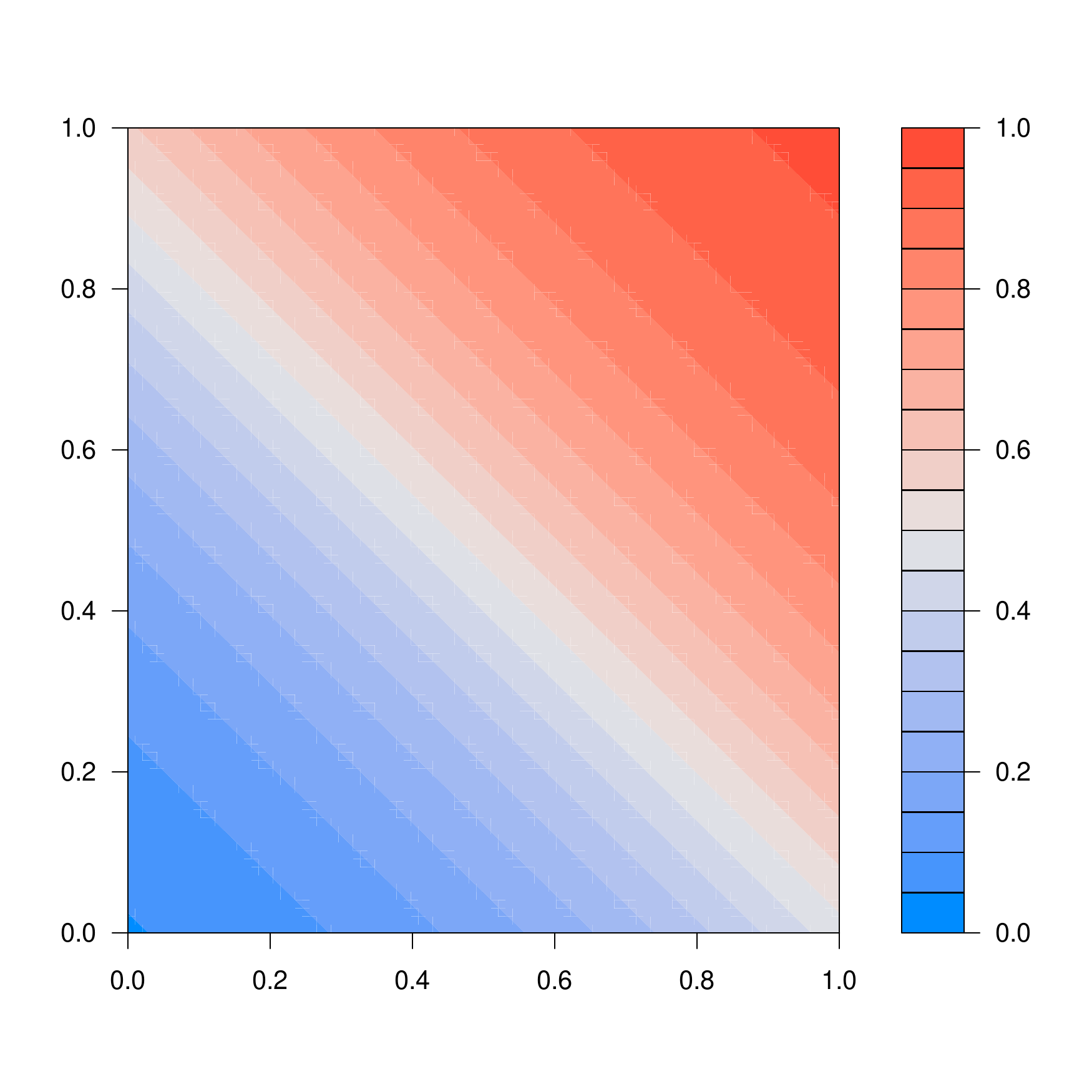} 
\subcaption{2-D Regression Surface}
\end{subfigure}
\caption{Probability(\textsf{declined}) depending on machine data and region.}
\label{fig:figRegDeclinedDataRegion}
}
\end{figure*}
}

\newcommand{\figRegDeclinedValueRegion}{
\begin{figure*}
{\centering
\begin{subfigure}[b]{0.49\textwidth}

\includegraphics[width=\maxwidth]{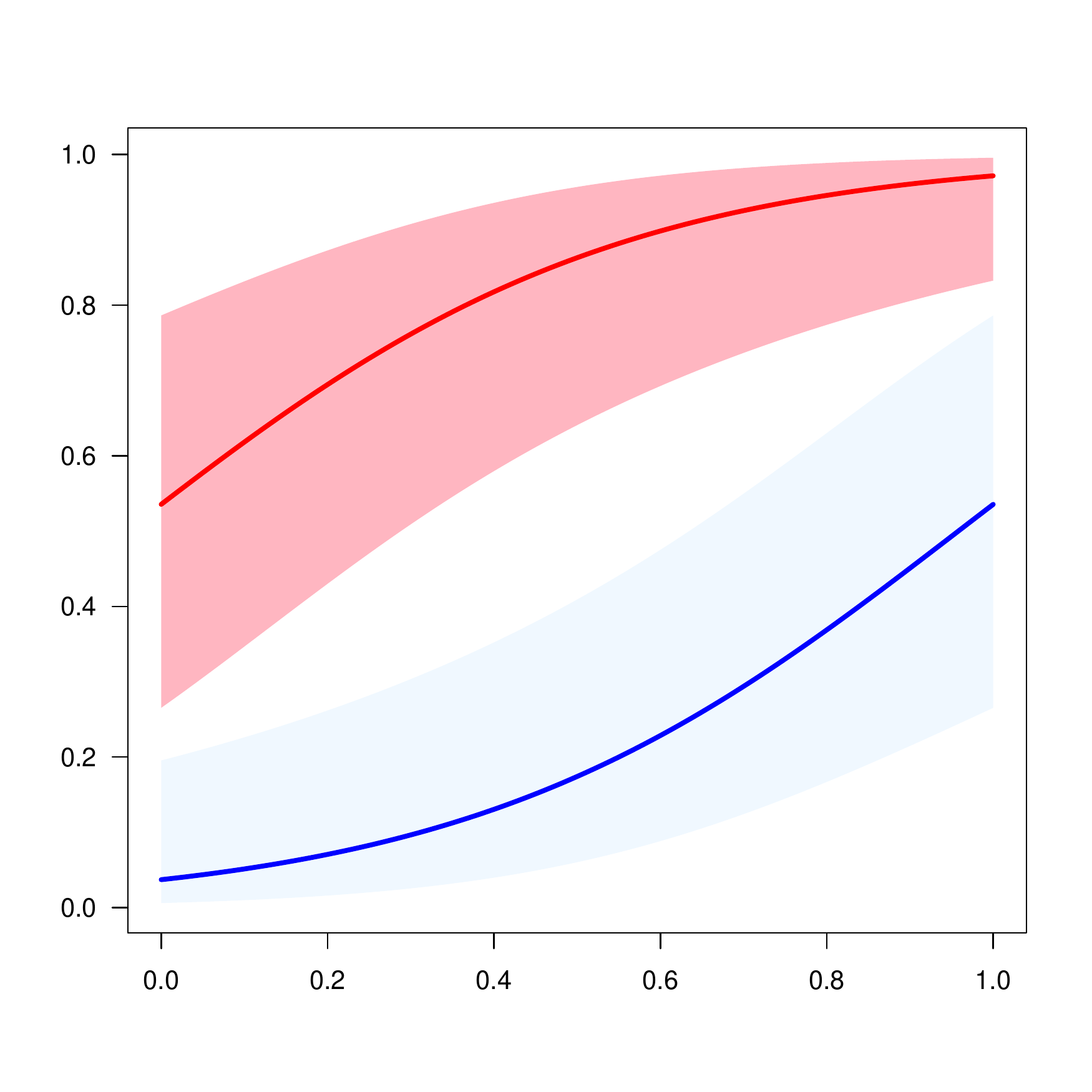} 
\subcaption{Overlay Plot \textsf{value} by \textsf{region}}
\end{subfigure}
\begin{subfigure}[b]{0.49\textwidth}

\includegraphics[width=\maxwidth]{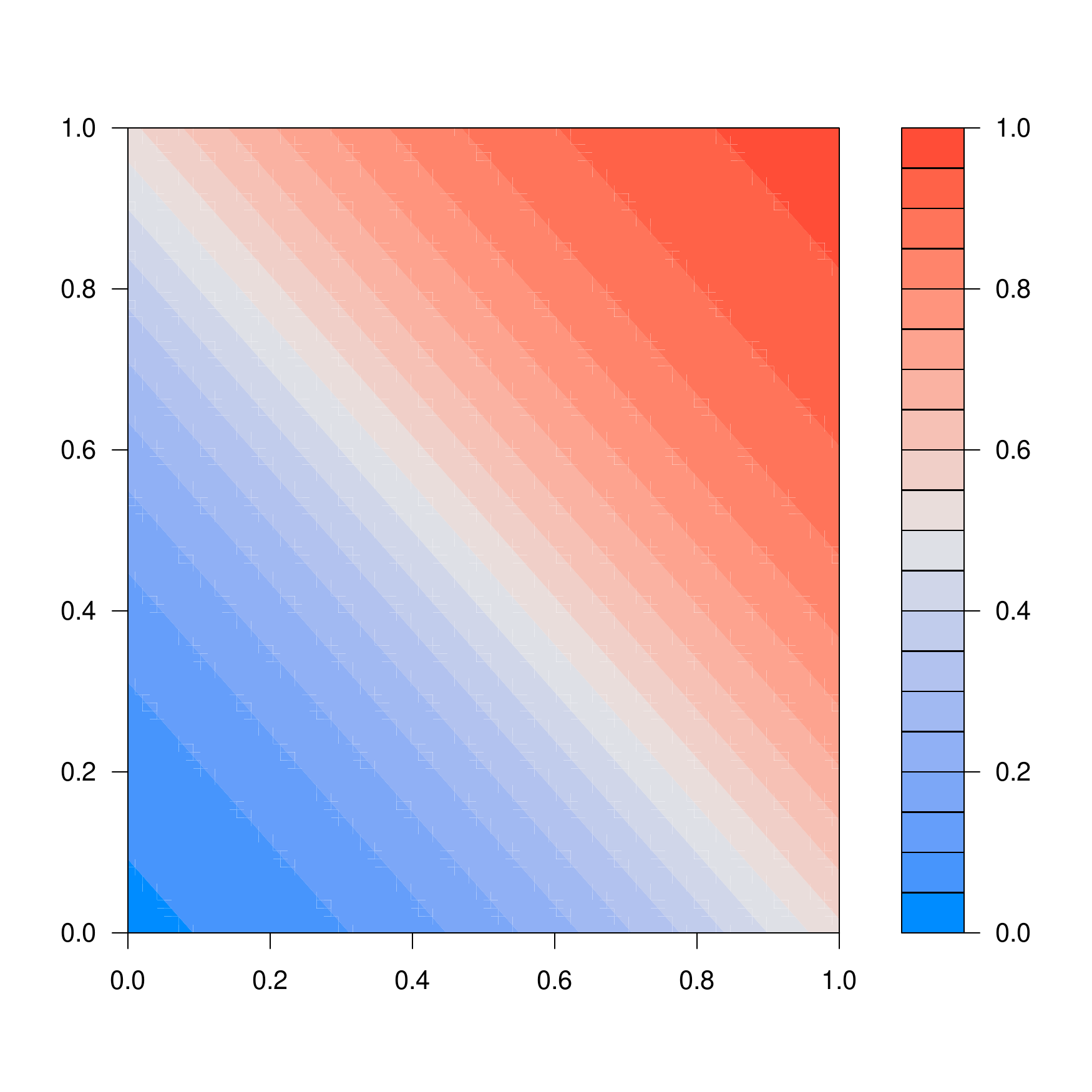} 
\subcaption{2-D Regression Surface}
\end{subfigure}
\caption{Overlay Plot: Probability(\textsf{declined}) depending on transaction value and region.}
\label{fig:figRegDeclinedValueRegion}
}
\end{figure*}
}

\newcommand{\figRegDeclinedDataValueRegion}{
\begin{figure*}[p]
{\centering
\begin{subfigure}[b]{0.49\textwidth}

\includegraphics[width=\maxwidth]{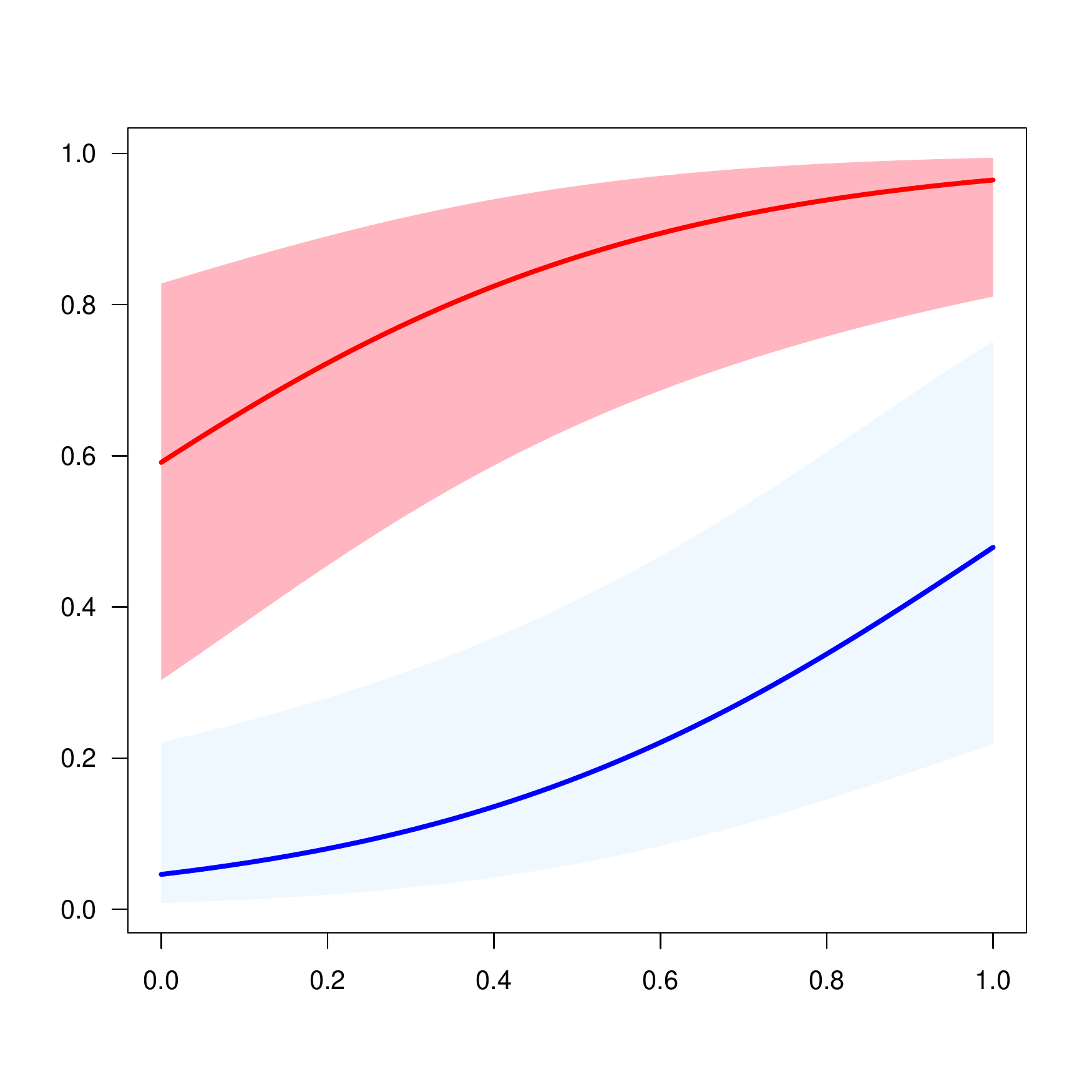} 
\vspace{-1cm}
\subcaption{\textsf{machinedata} by \textsf{region}}
\end{subfigure}
\begin{subfigure}[b]{0.49\textwidth}

\includegraphics[width=\maxwidth]{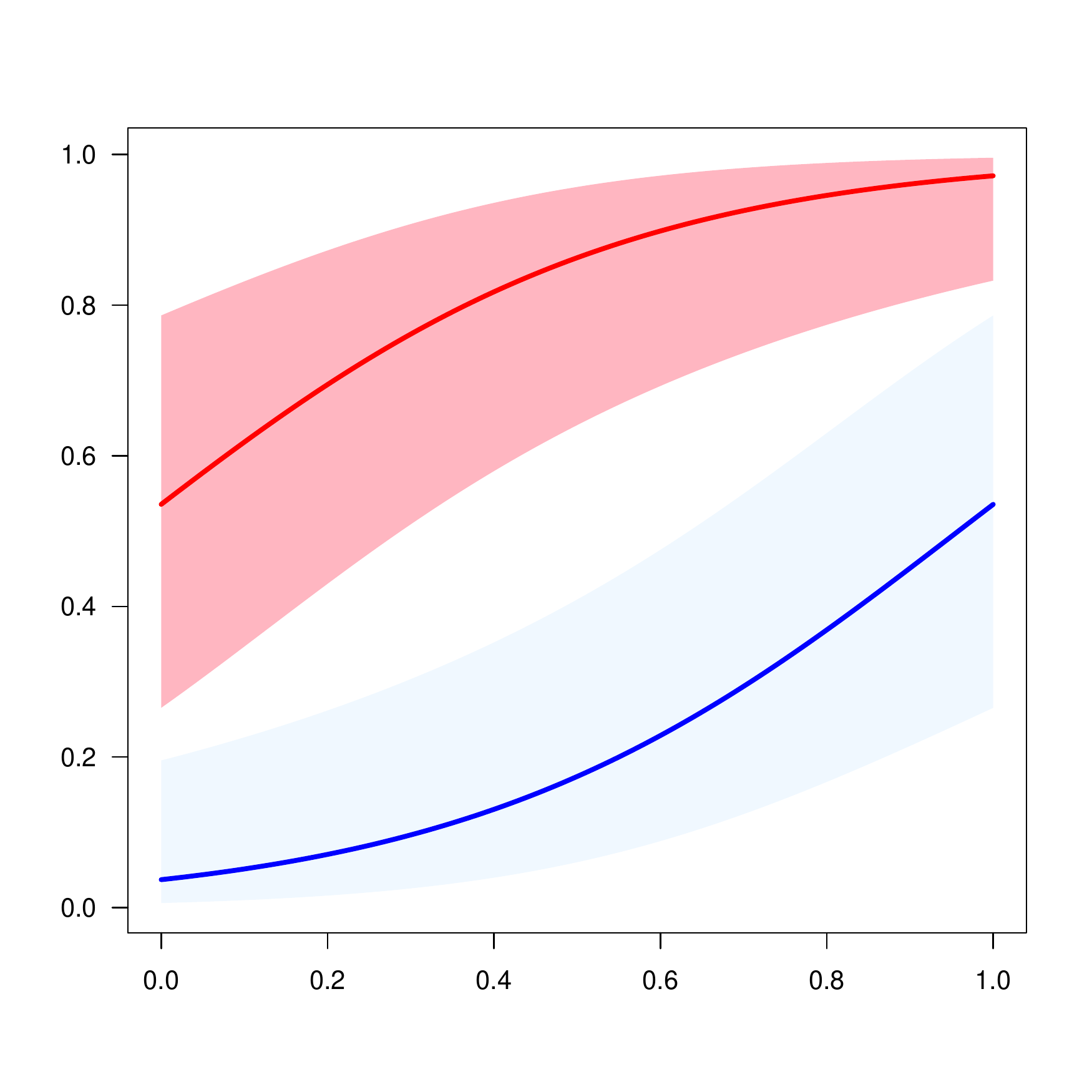} 
\vspace{-1cm}
\subcaption{\textsf{value} by \textsf{region}}
\end{subfigure}
\caption{Overlay Plot: Probability(\textsf{declined}) depending on [machine data \emph{or} transaction value] by region.
           (The blue line on the bottom shows the home region, the red line on the top the foreign region)}
\label{fig:RegDeclinedDataValueRegion}
}
\end{figure*}
}

\newcommand{\tabORChallenged}{
\begin{table*}[ht]
\centering
\caption{Odds Ratios: User Challenged} 
\label{tab:ORChallenged}
\begingroup\footnotesize
\begin{tabular}{rrrr}
  \hline
 & OR & LL & UL \\ 
  \hline
(Intercept) & 0.051 & 0.005 & 0.310 \\ 
  Machine.Data & 6.636 & 1.746 & 31.726 \\ 
  Value & 4.471 & 1.203 & 19.947 \\ 
  Region & 6.636 & 1.746 & 31.726 \\ 
  Website & 0.803 & 0.212 & 2.964 \\ 
  Card & 0.570 & 0.295 & 1.022 \\ 
   \hline
\end{tabular}
\endgroup
\end{table*}
}
\newcommand{\figORChallenged}{
\begin{figure}[htb]
{\centering
\vspace{-2.5cm}
\includegraphics[width=\maxwidth]{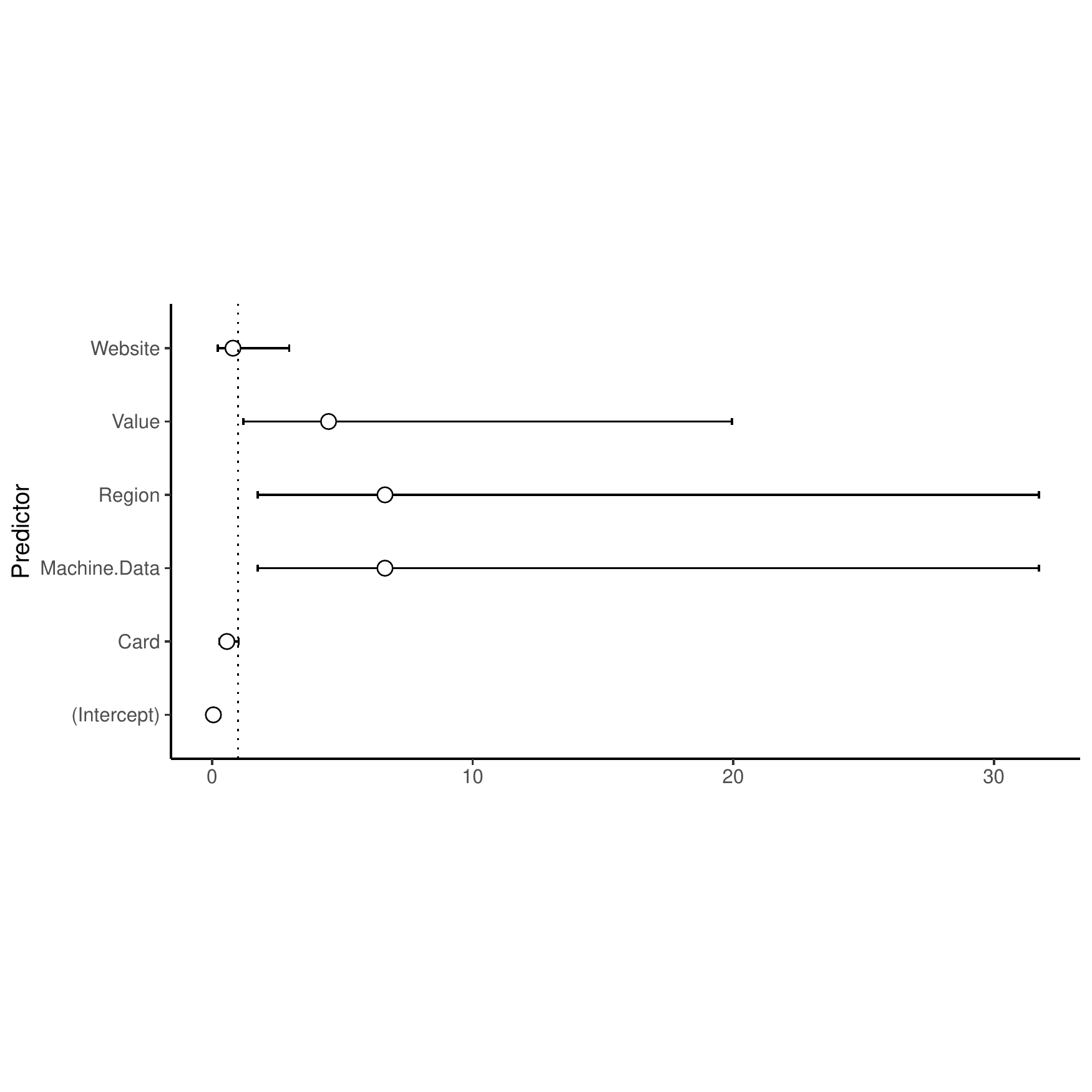} 
\vspace{-2.5cm}\caption{Odds Ratios Challenged, 95\% Confidence Intervals.}
\label{fig:OR.Challenged}
}
\end{figure}
}

\newcommand{\tabORDeclined}{
\begin{table*}[ht]
\centering
\caption{Odds Ratios: Transaction Declined} 
\label{tab:ORDeclined}
\begingroup\footnotesize
\begin{tabular}{rrrr}
  \hline
 & OR & LL & UL \\ 
  \hline
(Intercept) & 0.002 & 0.000 & 0.030 \\ 
  Machine.Data & 18.993 & 3.651 & 162.241 \\ 
  Value & 29.885 & 5.365 & 292.912 \\ 
  Region & 29.885 & 5.365 & 292.912 \\ 
  Website & 1.330 & 0.300 & 6.178 \\ 
  Card & 2.652 & 1.302 & 6.416 \\ 
   \hline
\end{tabular}
\endgroup
\end{table*}
}
\newcommand{\figORDeclined}{
\begin{figure}[htb]
{\centering
\vspace{-2.5cm}
\includegraphics[width=\maxwidth]{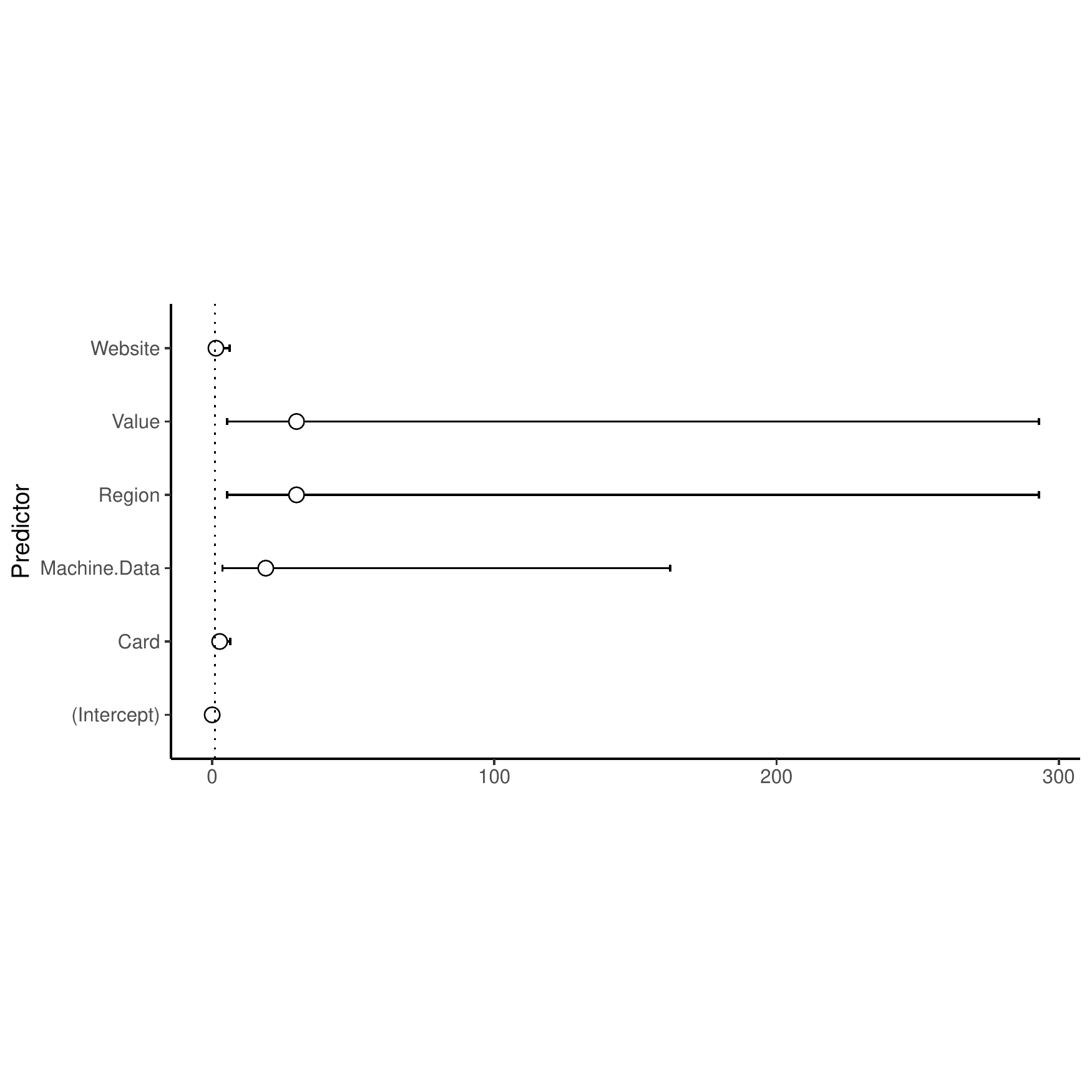} 
\vspace{-2.5cm}\caption{Odds Ratios Declined, 95\% Confidence Intervals.}
\label{fig:OR.Declined}
}
\end{figure}
}

\newcommand{\rocFigureSimple}{
\begin{figure}
{\centering

\includegraphics[width=\maxwidth]{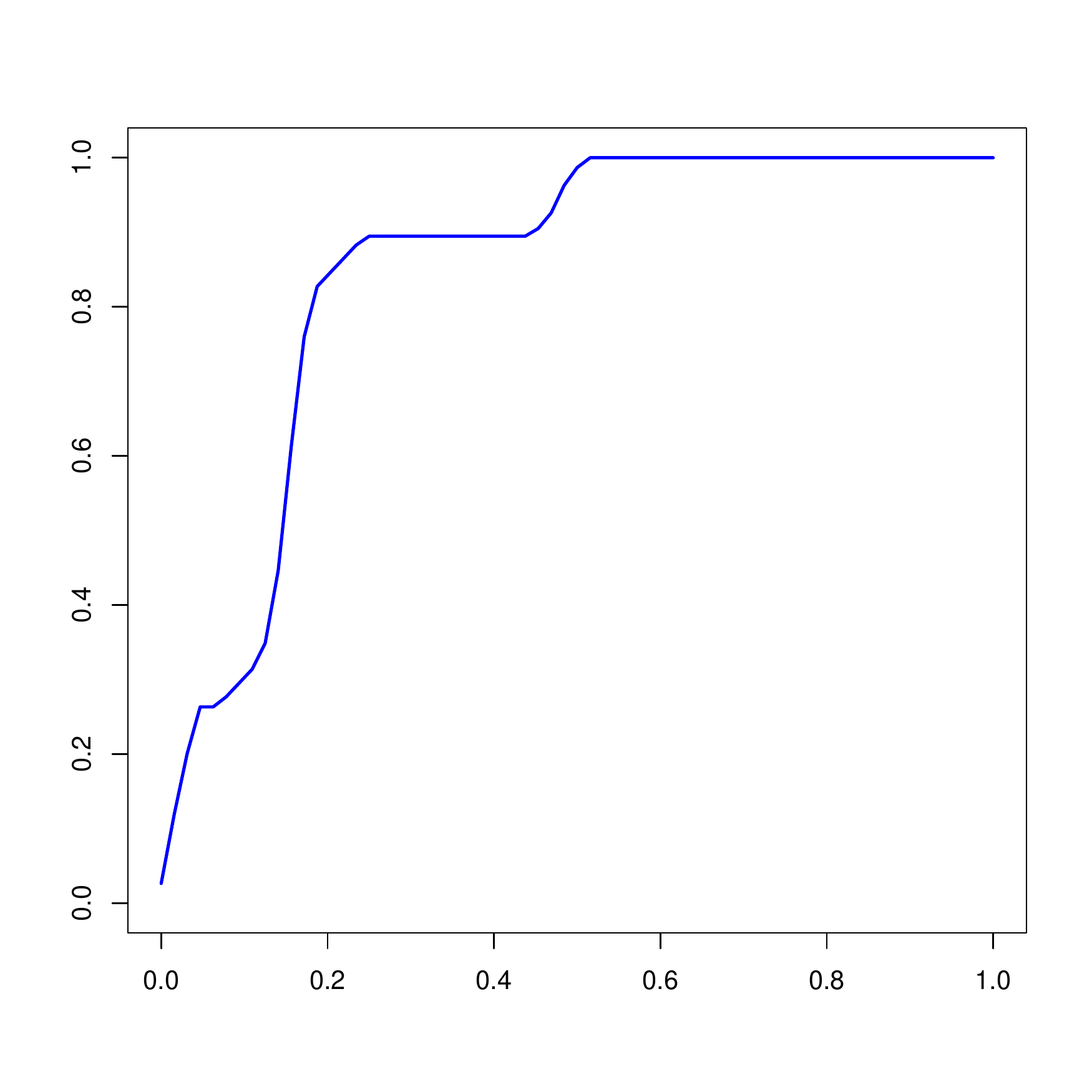} 
}
\end{figure}
}

\newcommand{\rocFigureCombined}{
\begin{figure}
{\centering

\includegraphics[width=\maxwidth]{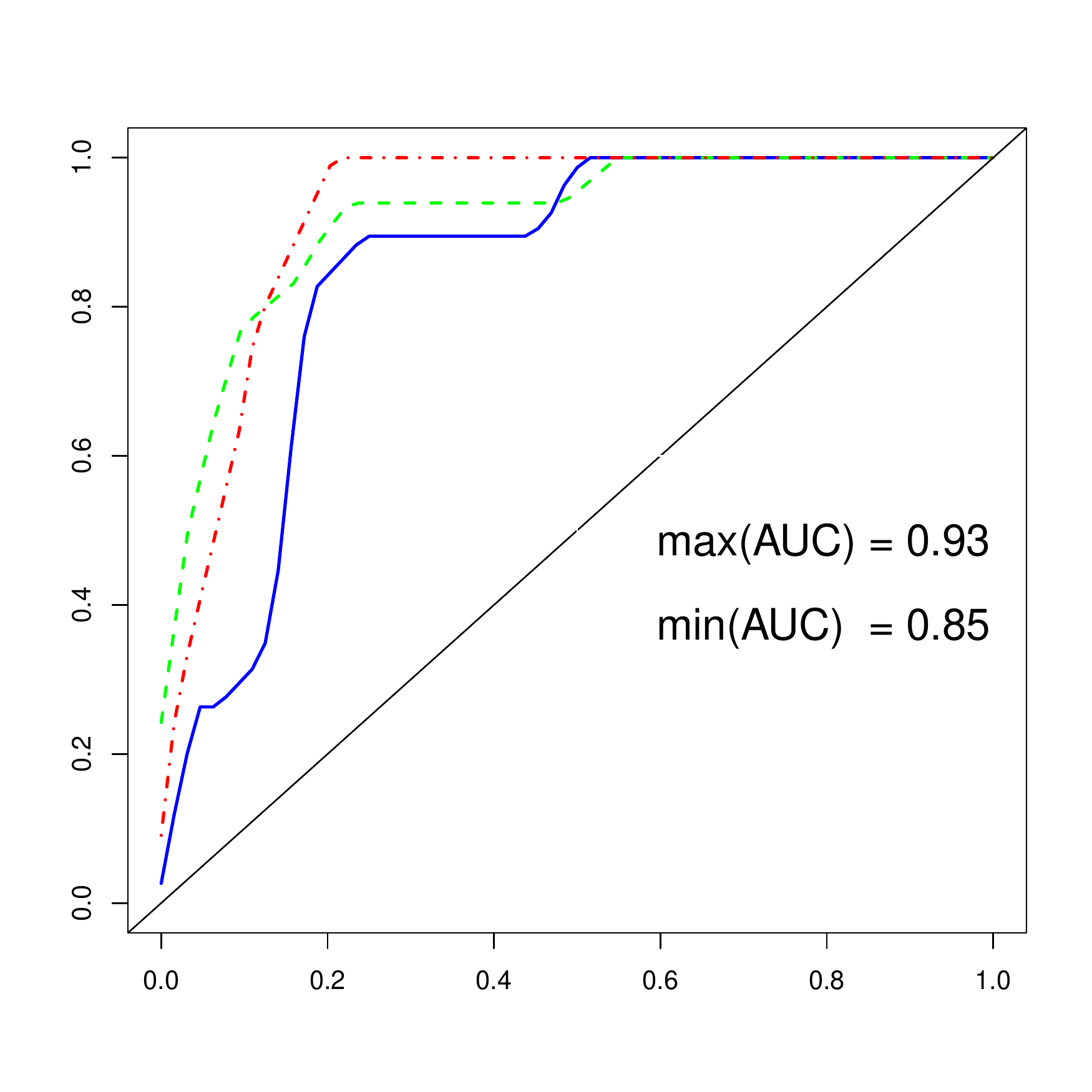} 
\caption{Receiver Operating Characteristic of the logistic regression classifiers.}
}
\end{figure}
}



\date{}

\begin{DocumentVersionSTAST}
\title{Investigation of 3-D Secure's Model for Fraud Detection}
\end{DocumentVersionSTAST}

\begin{DocumentVersionTR}
\title{Investigation of 3-D Secure's Model for Fraud Detection\thanks{Open Science Framework: \protect\url{https://osf.io/x6yfh}. This is the author's copy of this work. The definitive version was published in Proceedings of the 8th Workshop on Socio-Technical Aspects in Security and Trust (STAST'18), ACM Press, December 2018, pp. 1--11, \protect\url{https://doi.org/10.1145/3361331.3361334}.}}
\end{DocumentVersionTR}

\author{Mohammed Aamir Ali\\
School of Computing\\
Newcastle University, UK
\and
Thomas Gro{\ss}\\
School of Computing\\
Newcastle University, UK
\and
Aad van Moorsel\\
School of Computing\\
Newcastle University, UK}

\maketitle

\begin{abstract}
\noindent\textbf{Background.}
3-D Secure 2.0 (3DS 2.0) is an identity federation protocol authenticating the payment initiator for credit card transactions on the Web.

\noindent\textbf{Aim.}
We aim to quantify the impact of factors used by 3DS 2.0 in its fraud-detection decision making process.

\noindent\textbf{Method.}

We ran credit card transactions with two Web sites systematically manipulating the nominal IVs \textsf{machine\_data}, \textsf{value}, \textsf{region}, and \textsf{website}.
We measured whether the user was \textsf{challenged} with an authentication, whether the transaction was \textsf{declined}, and whether the card was \textsf{blocked} as nominal DVs.


\noindent\textbf{Results.}
While \textsf{website} and \textsf{card} largely did not show a significant impact on any outcome, \textsf{machine\_data}, \textsf{value} and \textsf{region} did. 

A change in \textsf{machine\_data}, \textsf{region} or \textsf{value} made it
5-7 times as likely to be challenged with password authentication. However, even in a foreign region with another factor being changed, the overall likelihood of being challenged only reached $60\%$.

When in the card's home region, a transaction will be rarely declined ($< 5\%$ in control, $40\%$ with one factor changed). 
However, in a region foreign to the card the system will more likely decline transactions anyway (about $60\%$) and any change in \textsf{machine\_data} or \textsf{value} will lead to a near-certain declined transaction.

The \textsf{region} was the only significant predictor for a card being blocked ($\const{OR}=3$).

\noindent\textbf{Conclusions.}
We found that the decisions to challenge the user with a password authentication, to decline a transaction and to block a card are governed by different 
weightings. 3DS 2.0 is most likely to decline transactions, especially in a foreign region. It is less likely to challenge users with password authentication, even if \textsf{machine\_data} or \textsf{value} are changed. 
\end{abstract}


\section{Introduction}
\label{s:introduction} 
Electronic commerce (e-commerce) is a mainstay of today's Internet, allowing users to buy or sell goods online. In payment systems terminology, e-commerce payments are known as Card Not Present (CNP) payments because the cardholder is not physically present at the merchant. CNP payment sales have shown a significant growth year-by-year. For example, in the UK, sales has been recorded a total of \pounds154 billion for 2017 \cite{Ecommerceeurope2017}. This is 18\% of increase in the online spending by customers when compared to year 2014 \cite{Ecommerceeurope2017}. 

The convenience of enabling purchases online comes at a price. The system is also prone to attract cyber offenders. Shown in \ref{fig:figFraud}, are the UK card payment fraud statistics from year 1998 to 2016. It can be seen from the figure that the payment industry is effective in mitigating card present types of payment card frauds. However, CNP payments fraud has reached its highest mark accounting for 70\% of the total card fraud, causing \pounds432.3 million loss exclusively on the UK issued cards \cite{FFAUK}. This development has called for a complex fraud detection system to be integrated with the protocol flows of the CNP payment system.

\pgfplotsset{compat=1.11}
\begin{DocumentVersionSTAST}
\begin{figure}
\centering
\begin{tikzpicture}
\begin{axis}[ bar width=0.2cm,
	width = 9cm,height=5cm,
    ymin=100, ymax=650,
    ytick = {50,100,200,300,...,650},
    ybar,
    enlargelimits=0.06,
    legend style={at={(0.5,-0.28)},
      anchor=north,legend columns=-1},
    symbolic x coords={1998,1999,2000,2001,2002,2003,2004,2005,2006,2007,2008, 2009, 2010, 2011, 2012, 2013, 2014, 2015, 2016},
    xtick=data,
    x tick label style={rotate=45,anchor=east},
    ]
\addplot[ybar, nodes near coords, every node near coord/.append style = {font=\tiny},  fill=black!9] coordinates {(1998,135) (1999, 188.4) (2000, 317.0) (2001, 411.5) (2002, 424.6) (2003, 420.4) (2004, 504.8) (2005, 439.4) (2006, 427.0) (2007, 535.2) (2008, 609.9) (2009, 440.1) (2010, 365.4) (2011, 340.9) (2012, 388.3) (2013, 450.4) (2014, 479.0) (2015, 567.5) (2016, 618.0) };
\addplot[draw=blue, thick,smooth] 
     coordinates {(1998,13.6) (1999, 29.3) (2000, 72.9) (2001, 95.7) (2002, 110.1) (2003, 122.1) (2004, 150.8) (2005, 183.2) (2006, 212.7) (2007, 290.5) (2008, 328.4) (2009, 266.4) (2010, 226.9) (2011, 220.9) (2012, 246.0) (2013, 301.0) (2014, 331.5) (2015, 398.2) (2016, 432.3)};
     \addplot[draw=red, thick,smooth] 
     coordinates {(1998,26.8) (1999, 50.3) (2000, 107.1) (2001, 160.4) (2002, 148.5) (2003, 110.6) (2004, 129.7) (2005, 96.8) (2006, 98.6) (2007, 144.3) (2008, 169.8) (2009, 80.9) (2010, 47.6) (2011, 36.1) (2012, 42.1) (2013, 43.4) (2014, 47.8) (2015, 45.3) (2016, 36.9)};
     \addplot[draw=black, thick,smooth] 
     coordinates {(1998,65.8) (1999, 79.7) (2000, 101.9) (2001, 114.0) (2002, 108.3) (2003, 112.4) (2004, 114.5) (2005, 89.0) (2006, 68.5) (2007, 56.2) (2008, 54.1) (2009, 47.7) (2010, 44.4) (2011, 50.1) (2012, 55.2) (2013, 58.9) (2014, 59.7) (2015, 74.1) (2016, 96.3)};
      \addplot[draw=gray, thick,smooth] 
     coordinates {(1998,34.9) (1999, 54.2) (2000, 103.5) (2001, 138.4) (2002, 130.2) (2003, 104.1) (2004, 92.5) (2005, 82.8) (2006, 117.1) (2007, 207.6) (2008, 230.1) (2009, 122.6) (2010, 93.9) (2011, 80) (2012, 101.6) (2013, 122) (2014, 150.3) (2015, 187.7) (2016, 200.1)};
\legend{Total,CNP,Counterfeit, Lost \& Stolen, Fraud Abroad}
\end{axis}
\end{tikzpicture}
\caption{UK Card Fraud by Type from year 1998 to 2016.}
\label{fig:figFraud}
\end{figure}
\end{DocumentVersionSTAST}

\begin{DocumentVersionTR}
\begin{figure*}
\centering
\begin{tikzpicture}
\begin{axis}[ bar width=0.2cm,
	width = 9cm,height=5cm,
    ymin=100, ymax=650,
    ytick = {50,100,200,300,...,650},
    ybar,
    enlargelimits=0.06,
    legend style={at={(0.5,-0.28)},
      anchor=north,legend columns=-1},
    symbolic x coords={1998,1999,2000,2001,2002,2003,2004,2005,2006,2007,2008, 2009, 2010, 2011, 2012, 2013, 2014, 2015, 2016},
    xtick=data,
    x tick label style={rotate=45,anchor=east},
    ]
\addplot[ybar, nodes near coords, every node near coord/.append style = {font=\tiny},  fill=black!9] coordinates {(1998,135) (1999, 188.4) (2000, 317.0) (2001, 411.5) (2002, 424.6) (2003, 420.4) (2004, 504.8) (2005, 439.4) (2006, 427.0) (2007, 535.2) (2008, 609.9) (2009, 440.1) (2010, 365.4) (2011, 340.9) (2012, 388.3) (2013, 450.4) (2014, 479.0) (2015, 567.5) (2016, 618.0) };
\addplot[draw=blue, thick,smooth] 
     coordinates {(1998,13.6) (1999, 29.3) (2000, 72.9) (2001, 95.7) (2002, 110.1) (2003, 122.1) (2004, 150.8) (2005, 183.2) (2006, 212.7) (2007, 290.5) (2008, 328.4) (2009, 266.4) (2010, 226.9) (2011, 220.9) (2012, 246.0) (2013, 301.0) (2014, 331.5) (2015, 398.2) (2016, 432.3)};
     \addplot[draw=red, thick,smooth] 
     coordinates {(1998,26.8) (1999, 50.3) (2000, 107.1) (2001, 160.4) (2002, 148.5) (2003, 110.6) (2004, 129.7) (2005, 96.8) (2006, 98.6) (2007, 144.3) (2008, 169.8) (2009, 80.9) (2010, 47.6) (2011, 36.1) (2012, 42.1) (2013, 43.4) (2014, 47.8) (2015, 45.3) (2016, 36.9)};
     \addplot[draw=black, thick,smooth] 
     coordinates {(1998,65.8) (1999, 79.7) (2000, 101.9) (2001, 114.0) (2002, 108.3) (2003, 112.4) (2004, 114.5) (2005, 89.0) (2006, 68.5) (2007, 56.2) (2008, 54.1) (2009, 47.7) (2010, 44.4) (2011, 50.1) (2012, 55.2) (2013, 58.9) (2014, 59.7) (2015, 74.1) (2016, 96.3)};
      \addplot[draw=gray, thick,smooth] 
     coordinates {(1998,34.9) (1999, 54.2) (2000, 103.5) (2001, 138.4) (2002, 130.2) (2003, 104.1) (2004, 92.5) (2005, 82.8) (2006, 117.1) (2007, 207.6) (2008, 230.1) (2009, 122.6) (2010, 93.9) (2011, 80) (2012, 101.6) (2013, 122) (2014, 150.3) (2015, 187.7) (2016, 200.1)};
\legend{Total,CNP,Counterfeit, Lost \& Stolen, Fraud Abroad}
\end{axis}
\end{tikzpicture}
\caption{UK Card Fraud by Type from year 1998 to 2016.}
\label{fig:figFraud}
\end{figure*}
\end{DocumentVersionTR}

\begin{figure*}[h]
\begin{center}
\tikzset{
  every picture/.append style={
    transform shape,
    scale=0.60
  }
}
\begin{sequencediagram}
\renewcommand\unitfactor{0.45}
  \newthread{customer}{:Payment Initiator}
  \newinst[4]{merchant}{:Merchant}
  \newinst[3.5]{acs}{Card Issuer}
  \newinst[4]{authorization}{:Authorization}
  \begin{call}{customer}{1. Pay / \texttt{https}}{merchant}{...X. Accept}
     \begin{call}{merchant}{2. Enable 3DS 2.0}{merchant}{activate3ds()}\end{call}
     \begin{mess}{merchant}{3. Connect ACS+POST[Tr.Num]}{customer}\end{mess}
     \begin{sdblock}{Tunnel(Customer,ACS)}{Frictionless Authentication Method}
       \begin{mess}{customer}{4. Connect ACS+POST[Transaction Number]+Cookies}{acs}\end{mess}
      \begin{call}{acs}{5. Load [dfp.js,SessionCookie]+Cookies}{customer}{7. POST[3DS Server Transaction ID, dfp.js (data)+Cookies]}
     \begin{call}{customer}{6. dfp.js}{customer}{}\end{call}
     \end{call}
     \end{sdblock}
     \begin{call}{merchant}{8. AReq}{acs}{10. AResp}
\begin{call}{acs}{9. Challenge}{acs}{}\end{call}
\end{call}

\begin{call}{merchant}{11. CReq}{acs}{19. CResp}
\begin{sdblock}{Loop}{Challenged Authentication Method}
\begin{call}{acs}{12. Challenge(OTP)}{customer}{14. Challenge Response(OTP)}
\begin{call}
{customer}{13. Enter OTP}{customer}{}
\end{call}
\end{call}

\begin{call}
{acs}{15. Determine Challenge}{acs}{Response}\end{call}\end{sdblock}
\begin{call}{acs}{16. RReq}{merchant}{18. RRes}
\begin{call}
{acs}{17. Receive and}{acs}{Log}
\end{call}
\end{call}
\end{call}
\begin{mess}{merchant}{20. Authorization}{authorization}\end{mess}
\end{call}
\end{sequencediagram}
\end{center}
\caption{Actions and parties involved in a 3DS 2.0 transaction process}
\label{fig:3ds}
\end{figure*}

  In general, the CNP payment system requires the payment initiator (customer) to enter their payment card information on the checkout page provided by the merchant website. The merchant collects the card information, combines it with the transactions information and forwards it to the card issuing bank for authorization. During the authorization process, the card issuer decides whether to approve or decline the transaction.
  
   Given that the payment card details are static  and are shared with every online merchant, there is a significant risk of the card data leaking and being used in fraudulent transactions. 
   Once the payment card details leaves the payment initiator's device, there is no guarantee that the card details are handled securely by the merchants. This is also reflected by recent attacks on Ticketmaster~\cite{BBCTM2018} and British Airways~\cite{BBCBA2018,FinancialTimesBA2018} and hundreds of websites where millions of card details were compromised~\cite{Worldpay}. Such systems lacks in the verification of the payment initiator as the valid owner of the card. Hence, the CNP payment system in itself is based on static card information and, as such, inherently insecure. 
  
  To protect the CNP payment system from fraud, VisaInc~\cite{VisaCo} introduced 3-D Secure 1.0 (3DS 1.0)~\cite{3DSecure} in 2001. 3-D Secure introduced the concept of payment initiator authentication for CNP payments. 3DS 1.0 redirects CNP transactions from each merchant website to the card issuer so that the payment initiator can be authenticated as the valid owner of the card. 

  With criticisms voiced on 3DS 1.0's registration and password authentication~\cite{Murdoch2010}, frictions in the checkout~\cite{Murdoch2010}, and the steady increase in CNP payment fraud, especially through phishing~\cite{KingRichard2009, Murdoch2010}, there was a need for a payment protocol upgrade. In 2016, EMVCo---a consortium of card payment networks---developed the 3D Secure 2.0 (3DS 2.0)~\cite{EMV3DS2.0} to address the requirements of stronger customer authentication yet maintaining the convenience requirements on a merchant checkout page. With 3DS 2.0, the card issuer performs fraud risk assessment for each transaction and authenticates the payment initiator with either of the two schemes: challenged and frictionless. Challenged authentication is designed for higher risks transactions and requires the payment initiator to authenticate him/herself with one-time pass codes sent by the card issuer to the payment initiator's registered device \cite{EMV3DS2.0}. Frictionless authentication is for purchases with lower risk of fraud and relies on the browser configuration details (hereafter referred to as browser fingerprint) extracted for the payment initiator device during the checkout process \cite{EMV3DS2.0}. At the same time, the decision making process of 3DS 2.0 to perform fraud risk assessment and the decision for challenged or frictionless authentication is shrouded from and often obscure to the consumers.
  
  In this paper, we quantify the impact of factors used by 3DS 2.0 in its fraud-detection decision making process. 
  That is, we aim at establishing to what extent a change in a factor changes the likelihood of a 3DS 2.0 decision outcome, e.g., whether the authentication is made challenged or frictionless.
  We run transactions with two Web sites manipulating Independent Variables (IV's) which includes machine\_data captured from a user Web browser (WB), transaction value, region and websites. To manipulate machine\_data, we set-up an HTTP proxy in the machine used to initiate transactions on 3DS 2.0 website. We measure whether the payment initiator was challenged with an authentication, whether the transaction was accepted with frictionless authentication or declined, and whether the card was blocked. We employ logistic regressions to quantify the change of likelihood observed from changes in the variables we have manipulated and, thereby, shine a light on the 3DS 2.0 fraud decision making process in the backend.
  
  This paper is organized as follows. Section~\ref{s:background} presents an overview of the 3DS 2.0 transaction process and provides an introduction into how the transaction risk assessment decisions are made by the card issuer. 
The paper follows an empirical-methods standard structure thereafter, describing the method first (Section~\ref{sec:method}), establishing the core results of the analysis without further interpretation (Section~\ref{sec:results}), and finally analyzing the results in a discussion (Section~\ref{sec:discussion}).
We draw attention to the logistic regression plots on pp.~\pageref{fig:RegPredChallenged} and~\pageref{fig:RegPredDeclined} as main tools to interpret the results.

\section{Overview of a 3DS 2.0 Transaction Process}
\label{s:background}
\begin{DocumentVersionTR}
 \begin{figure*}
 \centering
 \includegraphics[width=\textwidth]{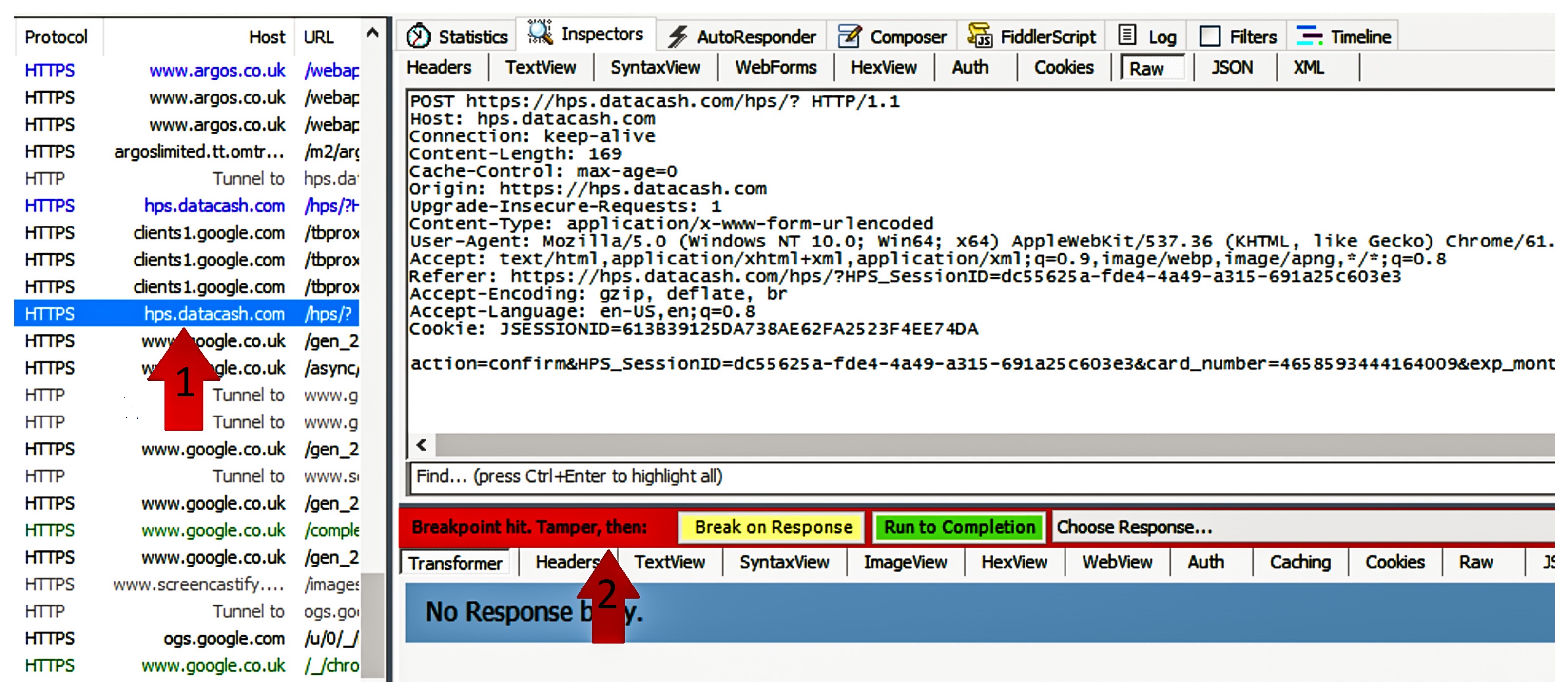}
 \caption{Screenshot of Fiddler proxy tool}
 \label{fig:proxy}
 \end{figure*}
\end{DocumentVersionTR}
Figure~\ref{fig:3ds} shows actions and parties involved in a 3DS 2.0 transaction process. The process starts with the payment initiator filling their payment card details on the checkout page provided by the merchant web site. When the ``Pay'' button is clicked, the merchant web server hosting the 3DS 2.0 plugin generates a unique transaction ID and connects the payment initiator's session to the card issuer. As shown in step 4, the card issuer connects to the payment initiator's Web Browser \textsc{(WB)} and sends device fingerprinting JavaScript (dfp.js) programmed to fetch browser and operating system details. The JavaScript mainly includes the following methods:

\begin{itemize}
    \item deviceprint\_browser(): This method extracts information about payment initiator's \textsc{(WB)} and operating system including: browser name, major and minor version, languages supported, languages installed, operating system name, operating system version, and operating system platform (Win32 or Win64).
    \item deviceprint\_display(): This method captures detailed screen information including colour depth, screen width, height, available height, buffer depth, and pixel depth.
    \item deviceprint\_software(): captures \textsc{(WB's)} plugins and their types. The method also has logic to extract browser's tracking and advertisement preferences as provided by DoNotTrack and Useofadblock.
    \item deviceprint\_java(): is used to test if the payment initiator browser supports Java or not.
    \item cookies(): is used to test if cookies are enabled by the user WB.
\end{itemize}
The information collected from the above methods is combined into a single string and is encoded into \textsf{base-64} plain text (as defined by the 3DS 2.0 protocol specifications \cite{EMV3DS2.0}) before being sent as a form element to the card issuer. It is likely that the card issuer uses IP address as an indicator to extract payment initiator machine location but it is captured differently.

In step 8, the merchant frames an Authentication Request \textsf{(AReq)} which is forwarded to the appropriate card issuer Access Control Server (ACS). The ACS manages 3DS 2.0 authentication request/response messages. The \textsf{AReq} contains card data provided by the payment initiator, merchant account information and other transaction related information. The card issuer collates the transaction information from the merchant and \textsf{WB} details provided by device fingerprinting scrips and performs fraud risk assessment \textsf{(FRA)} on the given transaction. Based on the outcome of \textsf{FRA}, the card issuer decides whether to challenge the payment initiator with a one-time pass codes or to authenticate the payment initiator with frictionless authentication. For the transaction shown in \ref{fig:3ds}, the card issuer decides to have challenged authentication.

In step 10, the issuer through the Authentication Response \textsf{(ARes)} message responds back to the merchant indicating that the challenge is required to further process the transaction.  For frictionless authentication, \textsf{ARes} will indicate a successful authentication.

The merchant initiates a Challenge Request \textsf{(CReq)} message and posts it to the card issuer. The issuer sends a challenge user interface \textsf{(UI)} to the payment initiator's \textsf{WB}. The \textsf{UI} is an interaction platform where the card issuer can interact with the payment initiator to obtain challenge response. At the point, the card issuer prompts challenge or \textsf{OTP} on the payment initiator's registered device (mobile phone for example).

The payment initiator enters the OTP on the 3DS 2.0 interface and upon successful authentication, the issuer  determines the payment initiator as  appropriate owner of the card and  formats the Results Request \textsf{(RReq)} message with a cryptographic hash which is forwarded to the merchant. The \textsf{RReq} and the hash is later used  by  the  Authorization network to verify the integrity of authentication messages. To acknowledge the receipt of the \textsf{RReq}, the merchant prepares the Results Response \textsf{(RRes)} and forwards it to the issuer. Finally, the issuer formats the Challenge Response \textsf{(CRes)} message  and  shuttles  it back to  the  merchant.  The \textsf{CRes}  indicates  the completion of challenged authentication. It is to be noted that the \textsf{CReq} and \textsf{CRes}  messages  are  only  applicable  to  challenged  3DS2.0 transaction.

\section{Aims}
\begin{researchquestion}[Impact of predictors on authentication outcomes]
Which factors impact the fraud-detection decisions with what magnitude of change in acceptance likelihood?
\end{researchquestion}
Table~\ref{tab:ops_auth} gives an overview of the operationalization of this research question.
As nominal independent variables (IV) we have \textsf{machine\_data}, \textsf{value}, \textsf{region}, and \textsf{website}.

As nominal dependent variables we consider whether the user was \textsf{challenged} with a password authentication, whether the transaction was \textsf{declined}\footnote{In the pre-registration of the experiment, the IV \textsf{declined} was called \textsf{transaction\_status}.} and whether the card was \textsf{blocked}.

Iterating over the independent variables $X \in \{$ \textsf{machine\_data}, \textsf{value}, \textsf{region}, \textsf{website} \} and the dependent variables $Y \in \{$ \textsf{challenged},  \textsf{declined}, \textsf{blocked} \}, we consider the following statistical hypotheses:
\begin{description}
  \item[Alternative Hypotheses.] $H_{1, X, Y}:$ The independent variable $X$ systematically impacts the likelihood of a change in the dependent variable $Y$.
  \item[Null Hypotheses.] $H_{0, X, Y}:$ The independent variable $X$ does not yield an impact on the likelihood of change in the dependent variable $Y$.
\end{description}
Note that we, thereby, investigate $5 \times 3$ relations with corresponding alternative and null hypotheses.
Our main interest lays in the IVs $\{$ \textsf{machine\_data}, \textsf{value}, \textsf{region} $\}$.

\paragraph{Logistic Regression Classifier.}
We use logistic regressions to establish the magnitude of impact on the likelihood on change in the response variable.

\begin{table*}[tb]
\centering
\footnotesize
\caption{Operationalization.}
\label{tab:ops_auth}
\begin{tabular}{ll}
\toprule
Variable          & Levels	   \\
\midrule
IV: \textsf{machine\_data} & \textsf{0} := intact \\
		& \textsf{1} := overwritten \\
IV: \textsf{value} & \textsf{0} := \textsf{low} (\textdollar{13}) \\
    & \textsf{1} := \textsf{high} (\textdollar{406})\\
IV: \textsf{region} & \textsf{0} := credit card home region (UK)\\
    & \textsf{1} := foreign region (Germany)\\
\midrule
DV: \textsf{challenged} &	\textsf{0} := \textsf{passed} (User passed without password authentication)\\
    & \textsf{1} := \textsf{challenged} (User was challenged with password authentication)\\
DV: \textsf{declined} & \textsf{0} := \textsf{accepted} (Transaction was accepted)\\
    & \textsf{1} := \textsf{declined} (Transaction was declined)\\
DV: \textsf{blocked} & \textsf{0} := \textsf{continued} (Credit card continued to be active)\\
    & \textsf{1} := \textsf{blocked} (Credit card was blocked by the bank)\\
\bottomrule
\end{tabular}
\end{table*}

\section{Experiments}
\label{s:experiments}
In a repeated-measures experiment, four different payment cards (including three Visa and a MasterCard) were used to make CNP payment transactions with two different Web sites that supports 3DS 2.0. 

Our experiments require manipulating \textsf{IV's} (\textsf{machine\_data, value, region}). Manipulating \textsf{machine\_data} requires the communication between the payment initiator's browser \textsf{WB}, the merchant and the card issuer to be intercepted (breakpoint in Fiddler terminology). We achieve this by placing an \textsf{HTTP} application proxy on the payment initiator's device (i.e., our own machine), as we describe below.

The \textsf{HTTP} application proxy tool that we use is Fiddler available at \cite{FiddlerTelerik}. Fiddler allows us to add breakpoint to alter the data before it is forwarded from \textsc{WB} to the communicating server. \processifversion{DocumentVersionTR}{In Figure~\ref{fig:proxy} the arrow labelled `1' adds a breakpoint when the user navigates to the payment URL on the merchant website `hps.datacash.com', and  the arrow labelled `2' points to where we edit the communication data to the merchant.} Using this platform, we are able 
\begin{inparaenum}[(i)]
\item to sniff the communication,
\item control the input to \textsf{WB}, and 
\item control the output from \textsf{WB}. 
\end{inparaenum}

To alter the \textsf{machine\_data}, we add two breakpoints. First, when the payment initiator click the `Pay' button on the merchant website. This is to modify the \textsf{HTTP} headers flowing from \textsf{WB} to the merchant. The second breakpoint we add is when card issuer connects to \textsf{WB} to fetch the browser fingerprint. This is to change the machine\_data. We alter the \textsf{HTTP} headers and the base-64 string of \textsf{WB} device fingerprint with that of recorded by Fiddler from a machine with different browser fingerprint. Collectively, we made 64 3DS 2.0 enabled transactions over two websites \textsf{argos.co.uk} and \textsf{bmstores.co.uk}, and over ranging transaction values of \$13 and \$406.

Collectively, we made 64 3DS 2.0 enabled transactions over two websites \textsf{argos.co.uk} and \textsf{bmstores.co.uk}, and over ranging transaction values of \$13 and \$406. To our experiments we studied the impact of fraud-detection decisions by performing transactions from two regions. Firstly by keeping the transaction region local to the country of where credit card is issued (UK) and region foreign to the credit card (Germany).

\section{Method}
\label{sec:method}
The study---its statistical hypotheses and analysis plan---have been preregistered at the Open Science Framework (OSF)\footnote{DOI 10.17605/OSF.IO/X6YFH; \url{https://osf.io/x6yfh/}} prior to any statistical analysis. 
Analyses, graphs and statistical reporting in this paper were computed directly from the data using the \textsf{R} package \textsf{knitr}. The OSF repository includes the dataset and its Datacite 4.0 meta-data description.

\subsection{Sampling}
In a repeated-measures experiment, four different payment cards (three Visa and a MasterCard) 
were used to make Card Not Present (CNP) payment transactions. 
We, thereby, sampled CNP payment transactions with 3DS-2.0-enabled home appliance Web sites (specifically: \textsf{argos.co.uk} and \textsf{bmstores.co.uk}).
The sampling frame was created by enumerating combinations of cards and Web sites, which were then exposed to different conditions.

\subsection{Procedure}
\begin{figure}
\centering
\includegraphics[height=1.5in, width=3in]{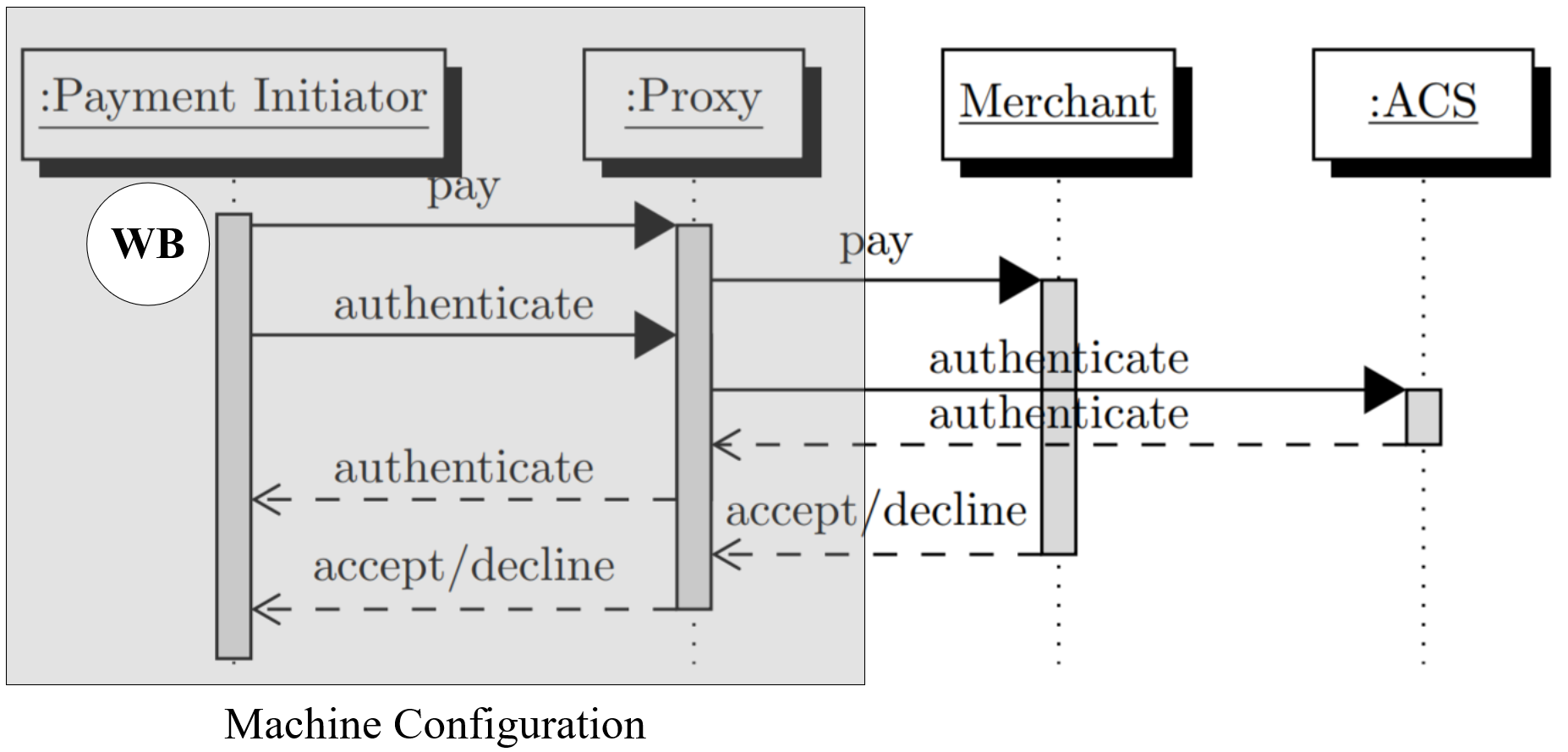}
\caption{Reverse engineering set-up, intercepting 3DS 2.0 transactions through a proxy.}
\label{fig:experiment}
\end{figure}
We manipulated the three \textsf{IVs} (\textsf{machine\_data, value, region}). To manipulate the \textsf{machine\_data}, we intercepted the communication between the payment initiator's browser \textsf{WB}, the merchant and the card issuer. We achieve this by placing Fiddler on the payment initiator's device (i.e., our own machine). Fiddler allows us to add breakpoint to alter the data before it is forwarded from \textsc{WB} to the communicating server. Using this platform, we are able 
\begin{inparaenum}[(1)] 
\item to sniff the communication, 
\item to control the input to \textsf{WB} and 
\item to control the output from \textsf{WB}. 
\end{inparaenum}

\subsubsection{Manipulation}
\paragraph{Machine Data}
To alter the \textsf{machine\_data}, we add two breakpoints. First, when the payment initiator click the `Pay' button on the merchant website. This is to modify the \textsf{HTTP} headers flowing from \textsf{WB} to the merchant. The second breakpoint we add is when card issuer connects to \textsf{WB} to fetch the browser fingerprint, as shown in Figure \ref{fig:experiment}. This is to change the \textsf{machine\_data}. We alter the \textsf{HTTP} headers and the base-64 string of \textsf{WB} device fingerprint with that of recorded by Fiddler from a machine with different browser fingerprint. 
 
\paragraph{Value}
To change the value of the transaction, we selected and purchased items that either cost \$13 or \$406.
 
 \paragraph{Region}
We kept the transaction either in a region local to the country of where credit card is issued (UK) or a region foreign to the credit card where the transactions were made from Germany. 
 
\subsubsection{Measurement}
We coded the outcomes of transactions on a nominal scale, either as `0' or `1', depending on whether the user could proceed with the transaction or was interrupted. This outcome was obtained from the response of the 3DS protocol to the browser.

We classified interruptions manually, based on 
\begin{inparaenum}[(1)]
  \item whether the user was challenged with a password authentication (\textsf{challenged}),
  \item whether the card transaction was declined (\textsf{declined}), and
  \item whether the payment card was blocked altogether (\textsf{blocked}).
\end{inparaenum}

\subsection{Ethics}
The experiment was run in accordance with the requirements of the institution's ethical review board and an ethics case signed off.

The payment cards used belonged to one of the experimenters, who exercised informed consent in volunteering the cards for the experiment.
The card holder was aware that repeated transactions as done in this experiment may impact future payment behavior of the card.

The card transactions were made by the card holder, and the relevant personal identifiable information not stored outside of the holder's control.

The transactions were done manually, over a longer timeframe, and restricted to at most 100 transactions, thereby, limiting the impact on the fraud detection efforts of providers, banks and 3DS.


\section{Results}
\label{sec:results}

The statistics were computed with a significance level of $\alpha = .05$. 
As a common approach, we conducted binomial logistic regressions with the dependent variables as response and a target model
including all independent variables. 

\subsection{Common Analysis Approach}
For each dependent variable, we created a logistic regression that is to quantify the change in likelihood caused by the different predictors.

\paragraph{Model Significance and Fit}
The first question is, whether a selected model is a valid and well-fitting model, at all.
We checked overall model significance with the Wald test, checked for significant higher-level interactions, and selected the final model using the Akaike Information Criterion corrected for small samples (\const{AIC_c}). Thereby, we ascertained that the selected
model contains substantial evidence vis-{\`a}-vis of the minimal-AIC model and substantiated its suitability with a goodness-of-fit check. 

While we aimed for a full model with all predictors as specified in the pre-registration, we checked that the model is actually defensible. While this was the case for the models on the user being challenged and on the transaction being declined, we found that for the card being blocked, there was not enough evidence to vouch for the full model. Here, we have selected a model only with the core predictors, as an alternative.

\paragraph{Impact of Predictors}
We computed odds ratios and $95\%$ confidence intervals thereon for effects of significant predictors.
Odds ratios are an effect size of choice for logistic regressions. They quantify the \emph{multiplicative} change in likelihood of the of the outcome, given a change in one predictor and everything else being equal. We may say, ``Other predictors held constant, a change in transaction value makes a rejection five times as likely.'' 

It is important to note that this multiplicative change is with respect to a baseline specific for each model. Hence, similar odds ratios in different models might lead to different absolute likelihoods of outcomes, given selected interventions.

\paragraph{Scenario Probabilities}
We also discuss the absolute likelihoods of outcomes for particular scenarios. Hence, then we factor in the odds ratios of active predictors and obtain the overall likelihood in that situation. In this case, we consider combinations of predictors being manipulated and offer a likelihood estimate for the outcome. Here we may say that ``In a foreign region, the transaction is $99\%$ likely to be declined if the machine data is faulty.''

\paragraph{Model Evaluation}
Finally, we evaluated each model with regression diagnostics as well as accuracy (prediction vs. observation).
We computed repeated 10-fold cross-validations with the \textsf{R} package \textsf{caret} (with 10 repetitions).

\processifversion{DocumentVersionSTAST}{We report the results of the model evaluation in Appendix~\ref{sec:model_eval}.}

\subsection{Logistic Regression: User Challenged}
We computed a binomial logistic regression to test whether the independent variables impact the likelihood of \textsf{challenged} as response. \processifversion{DocumentVersionSTAST}{Appendix~\ref{sec:model_eval_challenged} contains details of the model selection and evaluation.}

\begin{DocumentVersionTR}
\paragraph{Model Selection}
The model with all five predictors was statistically significant, $\chi^2(5) = 21.593$, $p < .001$. We checked for second-level and third-level interactions wrt. the manipulated independent variables and found none.

This model yields an Akaike Information Criterion corrected for small sample sizes of $\const{AIC_c} = 69.73$.
Compared to the best model only including the predictors with the greatest residual drop (\textsf{machine\_data}, \textsf{value}, and \textsf{region}),
this fitted model experiences a small enough information loss ($\Delta = 1.14$) to be classified as having substantial support.

\paragraph{Goodness of Fit}
We computed a likelihood-ratio test to compare the full model selected against the one with minimal \const{AIC_c} (and fewer predictors).
We failed to reject the null hypothesis that the reduced model is true, and, hence, keep the full model. We report the goodness-of-fit in different variants of Pseudo-$R^2$ in Table~\ref{tab:regTableChallengedRefined}. McFadden's $R^2 = .28$.
\end{DocumentVersionTR}

\subsubsection{Fitted Model}

\paragraph{Predictors and Odds Ratios}
Table~\ref{tab:regTableChallengedRefined} offers an overview estimates of the selected model.

\tabRegTableChallengedRefined

There was a statistically significant positive impact of overwriting the machine data on the user being challenged with a password, $z = 2.6$, $p = .009$, $\const{OR} = 6.64$, 95\% CI $[1.7, 31.7]$. Everything else being equal, a user whose machine data is corrupted is $6.6$ times as likely to be challenged with a password authentication.

A change of the value of a transaction from low to high had a statistically significant effect on the user being challenged, $z = 2.12$, $p = .034$, $\const{OR} = 4.47$, 95\% CI $[1.2, 19.9]$. The increase in value from \textdollar{13} to \textdollar{406} made it $4.5$ times as likely to be challenged.

The region being changed to a foreign country had a statistically significant impact on being challenged with a password authentication, $z = 2.6$, $p = .009$, $\const{OR} = 6.64$, 95\% CI $[1.7, 31.7]$. A change in region to Germany made it $6.6$ times as likely.

Given these results we reject the null hypotheses $H_{0, X, \const{challenged}}$ for $X \in  \{$ \textsf{machine\_data}, \textsf{value}, \textsf{region} $\}$.

\paragraph{Scenario Probabilities}
We display the response plots of the different predictor variables in Fig.~\ref{fig:RegPredChallenged} on p.~\pageref{fig:RegPredChallenged}.
Note that given the similar odds ratios of the predictors, we expect the response and overlay plots below to look rather similar to one another.

In Fig.~\ref{fig:RegChallengedDataValueRegion}, we overlay by region the likelihoods of changing machine data or value, respectively.
Overall, we observe that the probability of being challenged while being in home region of the card is less than $5\%$ when neither machine data nor value are manipulated. Should either of the two predictors be changed (machine data overwritten or the value high), the probability of being challenged is less than $20\%$.

If the card does a transaction from the foreign region, the situation is quite different.
Here, the card will challenge the user at probability of $20\%$ or $25\%$ even if machine data are intact and the value low.
Should either of the two variables be manipulated, the user will be challenged at a probability of around $60\%$.

\figRegPredChallenged

\figRegChallengedDataValueRegion

\subsubsection{Model Evaluation}
\begin{DocumentVersionTR}
\paragraph{Diagnostics}
There were two cases with large residuals, but DFbetas were well below $.5$. There were no cases with large leverage.
Assessing for multicollinearity, we found the Variance Inflation Factors (VIFs) all close to $1$, with a mean VIF of $1.09$.
\end{DocumentVersionTR}

\paragraph{Performance}
Having computed a observation-vs.-prediction classification, we found that the regression had an accuracy of $73\%$, Hosmer-Lemeshow not rejecting the fit, \const{HL_C} $\chi^2(8) = 10.77$, $p = .215$.

\paragraph{Cross-Validation}
We computed a repeated $10$-fold cross-validation on the same dataset. 
This means, that the dataset was partitioned into 10 parts, that the model was then re-computed using 9 parts as training data (with $N_T = 57\pm 1$), and used to predict the observations of the $10$-th part.

The cross-validation yielded an accuracy of $71\%$, 95\% CI $[67\%, 74\%]$. With a Cohen's $\kappa = .25$, we consider the cross-validation accuracy as low.

\subsection{Logistic Regression: Transaction Declined}
We computed a binomial logistic regression with the independent variables as predictors and the transaction being declined as response variable. \processifversion{DocumentVersionSTAST}{We report on the model evaluation in Appendix~\ref{sec:model_eval_declined}.}

\begin{DocumentVersionTR}
\paragraph{Model Selection}
The model including all predictors was statistically significant, $\chi^2(5) = 44.409$, $p < .001$. Second-level and third-level interactions between manipulated variables were not statistically significant.

This model comes with a corrected Akaike Information Criterion $\const{AIC_c} = 57.72$.
Compared to the min-$\const{AIC_c}$ model with the predictors \textsf{machine\_data}, \textsf{value}, \textsf{region}, and \textsf{card},
the fitted model has an information loss of $\Delta = 2.3$. 
This is which past the threshold of substantial support, but still considered good evidence.

\paragraph{Goodness of Fit}
Comparing the min-$\const{AIC_c}$ model with the chosen model on goodness-of-fit, we find that a likelihood-ratio test does not reject the null hypothesis.
We report the goodness-of-fit in Pseudo-$R^2$ in Table~\ref{tab:regTableDeclinedRefined}. McFadden's $R^2 = .50$.
\end{DocumentVersionTR}

\subsubsection{Fitted Model}

\paragraph{Predictors and Odds Ratios}
We are offering an overview of all estimates in the regression Table~\ref{tab:regTableDeclinedRefined}.

Overwriting the machine data has a statistically significant impact on the transaction being declined, $z = 3.12$, $p = .002$, $\const{OR} = 18.99$, 95\% CI $[3.7, 162.2]$. Everything else being equal, overwriting the machine data made it $19$ times as likely to get the transaction declined.

There was a statistically significant effect of the value of the transaction on it being declined, $z = 3.41$, $p < .001$, $\const{OR} = 29.88$, 95\% CI $[5.4, 292.9]$. Other predictors held constant, increasing the value to \textdollar{406} made it $29.9$ times as likely to have transaction declined.

The region had a statistically significant effent on the transaction being declined, $z = 3.41$, $p < .001$, $\const{OR} = 29.88$, 95\% CI $[5.4, 292.9]$.
Other predictors constant, a change to the foreign region (Germany) made it $29.9$ times as likely to have transaction declined.

We thereby reject the null hypotheses $H_{0, X, \const{declined}}$ for $X \in  \{$ \textsf{machine\_data}, \textsf{value}, \textsf{region} $\}$.

In addition to these predictors, the card used also had a statistically significant effect on the transaction being declined.

\tabRegTableDeclinedRefined

\paragraph{Scenario Probabilities}
We are giving an overview of regression (response and overlay) graphs for the transaction-declined regression in Figures~\ref{fig:RegPredDeclined} and~\ref{fig:RegDeclinedDataValueRegion} on p.~\pageref{fig:RegPredDeclined}.

As expected, the response graphs shown in Fig.~\ref{fig:RegPredDeclined} are similar due to the similar odds ratios of the predictors in question.

\figRegPredDeclined

\figRegDeclinedDataValueRegion

We consider the overlay of [machine data or transaction value] by region in Fig.~\ref{fig:RegDeclinedDataValueRegion}.

In the home region, we observe that the probability to get a transaction declined in below $5\%$, if machine data and value stay at control level.
If either of them are changed to machine data being overwritten or the value increased, we expect a probability of about $50\%$ to get the transaction declined.

Once the user requests a transaction from the foreign region, the probabilities are not in the user's favor.
Everything else at control level, we expect a probability of $50\%$ to $60\%$ of the transaction being declined.
If either the machine data is overwritten or value is increased, it is almost certain for the transaction to be declined.

\subsubsection{Model Evaluation}
\begin{DocumentVersionTR}
\paragraph{Diagnostics}
There were four cases with large residuals, yet DFbetas shown to be below $1$.
There were 12 cases with leverage just touching twice the average leverage, however the DFbetas are consistently less than $1$ and Cook's distance less than $0.1$.
The VIFs are smaller than $2$, where the mean VIF is $1.49$.
\end{DocumentVersionTR}

\paragraph{Performance}
We have a classification accuracy of $83\%$, Hosmer-Lemeshow not rejecting the fit, \const{HL_C} $\chi^2(8) = 6.96$, $p = .540$.

\paragraph{Cross-Validation}

The repeated $10$-fold cross-validation showed an accuracy of $78\%$, 95\% CI $[75\%, 82\%]$. The model offers reasonably accuracte predictions (Cohen's $\kappa = .57$), with accuracy statistically significantly greater than the no-information rate, $p < .001$.

\subsection{Logistic Regression: Card Blocked}
We established a binomial logistic regression on the impact of predictors \textsf{machine\_data}, \textsf{value} and \textsf{region} on the credit card being blocked. \processifversion{DocumentVersionSTAST}{We offer details on model selection and evaluation in Appendix~\ref{sec:model_eval_blocked}.}

\begin{DocumentVersionTR}
\paragraph{Model Selection}
We have a scenario in which the model including only \textsf{value} and \textsf{region} has the minimal $\const{AIC_c} = 40.39$.

The full model including the three other predictors ($\const{AIC_c} = 44.71$) yields an information loss of $\Delta = 4.32$, having considerably less support. Given the data, this model only carries a likelihood of $1\%$. 

Even though the likelihood-ratio test does not reject the null hypothesis, we consider the model with \textsf{value}, \textsf{region} and\textsf{machine\_data} as robust alternative ($\const{AIC_c} = 41.05$). In comparison with the minimal-$\const{AIC_c}$ model, we have an information loss of $\Delta = 0.67$, yielding substantial evidence. Hence, we select this model.

\paragraph{Goodness of Fit}
We evaluate a likelihood-ratio test to check the goodness-of-fit of the min-$\const{AIC_c}$ model vis-{\`a}-vis the chosen model.
It did not reject the null hypothesis.
Table~\ref{tab:regTableBlockedRefinedInt} contains customary Pseudo-$R^2$ estimates. McFadden's $R^2 = .45$.
\end{DocumentVersionTR}

\subsubsection{Fitted Model}

\paragraph{Predictors and Odds Ratios}
We offer an overview of the predictor estimates and $p$-values in Table~\ref{tab:regTableBlockedRefinedInt}.

\tabRegTableBlockedRefinedInt

The only predictor statistically significantly impacting the likelihood of the card being blocked was the region, $z = 1.23$, $p = .217$, $\const{OR} = 2.99$, 95\% CI $[0.6, 19.8]$. A change from the card's home region (UK) to the foreign region (Germany) made it $3$ times more likely to get the card blocked.

We thereby failed to reject the null hypotheses $H_{0, X, \const{blocked}}$ for $X \in  \{$ \textsf{machine\_data}, \textsf{value}, \textsf{region} $\}$.

\subsubsection{Model Evaluation}
\begin{DocumentVersionTR}
\paragraph{Diagnostics}
There was one case with high residuals, but DFbetas smaller than $1$.
There were $9$ cases with a leverage past the double-mean-leverage threshold.
Inspecting DFbetas, we find them to be below $1$, and inspecting the cooks distance, we find it below $0.2$ max.
The VIF was consistently close to 1, with a mean VIF of $1.04$.
\end{DocumentVersionTR}

\paragraph{Performance}
The classification accuracy was $73\%$, Hosmer-Lemeshow not rejecting the fit, \const{HL_C} $\chi^2(8) = 1.78$, $p = .987$.

\paragraph{Cross-Validation}
The repeated $10$-fold cross-validation yielded an accuracy of $84\%$, 95\% CI $[81\%, 87\%]$, Cohen's $\kappa = .40$.

\subsection{Overall Model Properties}
The three selected models stay valid with each $p < .001$ under Bonferroni-Holm correction for multiple comparisons.
The selected models show a classification accuracy of $73\%-83\%$, with a passable fit.
At the same time, the cross-validation accuracy was low to medium, which makes the selected models less useful as predictive classifiers for other datasets.


\section{Discussion}
\label{sec:discussion}
The discussion is best seen in context of the likelihood plots of Figures~\ref{fig:RegPredChallenged} and~\ref{fig:RegPredDeclined} on pp.~\pageref{fig:RegPredChallenged} and~\pageref{fig:RegPredDeclined}.

\subsection{The overall likelihoods of outcomes differ characteristically.}
From the quantification on likelihoods obtained from the logistic regressions, we can observe a consistent order of likelihoods.
It was most likely for a transaction to be rejected especially in a foreign region.
It seems that 3DS is taking no chances in the case of either the machine data being corrupt or the value being too high: the that transactions are declined is all but certain.

It is noteworthy that the likelihood to decline transactions was consistently higher in foreign and home regions alike than the likelihood to challenge the user with a password authentication. There seems to be a prioritization of user convenience in the sense creating less interruptions in payment flow overall.

Of the three outcomes considered, the card being blocked had the lowest effect size (odds ratio), that is, 3DS seems least likely to have a card blocked as \emph{ultima ratio}. Of course, this makes sense given the hassle for consumers and banks alike to get a card unblocked or a new card issued.

\subsection{The three independent variables have an effect in the same order of magnitude.}
For being challenged and the transaction declined, we find that the three interventions investigated (overwriting machine data, changing to a foreign region, or increasing the transaction value) all yielded an impact on the respective outcome with a change in likelihood roughly in the same order of magnitude. Hence, we conclude that 3DS takes into account all three variables in its decision making process and that the variables are roughly equally weighted. 

It is important to note, however, that the variables are not KO criteria: If something is amiss in only one of those variables, the outcome will just be biased towards the user being challenged or the transaction being declined. However, only if two variables come together in a deviation from the norm (machine data intact, home region, value relatively low) then the likelihood of a 3DS intervention is predicted to be more than $50\%$.

\subsection{The impact of the region is consistently strong.}
Having said that the three variables seem to have an impact of equal order of magnitude, it seems that the region still ranks first among them, consistently being in the first effect-size rank. This becomes especially apparent when looking at the ``by region'' plots presented in Figures~\ref{fig:RegChallengedDataValueRegion} and~\ref{fig:RegDeclinedDataValueRegion} on pp.~\pageref{fig:RegChallengedDataValueRegion} and~\pageref{fig:RegDeclinedDataValueRegion}.

Here, we see that the change from home to foreign region impacts the likelihood of a negative outcome more strongly than the combined changes in likelihood caused by problems with machine data or transaction value.

\subsection{The impact of the card used seemed consistently weak.}
While we used four different payment cards from different providers in the experiment, we found consistently low effect sizes on the impact of the card used.

\subsection{Limitations}
\subsubsection{Generalizability}
We are the first to state that the generalizability of this experiment is somewhat limited.
In terms of experiment design, this is rooted in a small number of credit cards, card providers and merchant sites being evaluated. Furthermore, the different payment cards used were linked to a single card holder. 

While the card being used generally was linked to a comparatively low effect size, the experiment was thereby not prepared to discern whether 3DS and payment card institutions \emph{personalize} their responses to the card holder.

To gain a quantification of the impact of different card holder profiles on the outcomes, one would need a wide range of participants with different credit card histories, which was beyond the scope of this study.

While such future research might yield interesting results, it comes with ethical caveats that participants might expose their credit card accounts with a host of failed transactions, which in turn could impact the future behavior of their payment cards.

\subsubsection{Sample Size \& Power}
Operating on live credit cards owned by real people, 
we saw a need to exercise restraint how many transaction we would run.

We computed the logistic regressions with a sample of $N=64$ and
a maximal number of predictors $k=5$.

An \emph{a priori} power analysis with \textsf{G*Power} based on a presumed $H_1$ probability of $50\%$ and 
a presumed $H_0$ probability of $20\%$ for the impact of one predictor (assuming the others to have $R^2 = .3$), 
highlighted a need of a minimal sample size of $\hat{N} = 55$ to reach $80\%$ power.

We are aware that we are operating below the rule-of-thumbs limits of sample sizes used for binomial logistic regressions.
We accepted that we accepted that in terms of sensitivity, we could only detect effect of $\const{OR}\geq 4$ at $80\%$ power.
To be prudent operating at this small a sample size, we used the corrected Akaike Information Criterion ($\const{AIC_c}$)
for the model selection and profile-likelihood limits for the interval estimation on odds ratios 
(both said to be superior for small sample sizes).

\section{Conclusion}
In this paper, we presented the first attempt to quantify back-end decision making process of 3-D Secure (3DS).
Considering the 3-D Secure decisions as probabilistic, we have employed an empirical experiment to evaluate to what extent different deviations from the norm (overwriting machine data, leaving a payment card's home region, or increasing the value of the transaction) change the likelihood of a ``negative'' outcome for the user.

To the best of our knowledge, we are the first to employ logistic regression to quantify the changes in likelihood in the outcomes of the 3DS decision, that is, whether the user is challenged with a password authentication, whether the transaction is declined, or whether the card is blocked altogether.

Overall, we observed that the likelihood of the different outcomes follow different distributions, transactions declined being the most likely, card blocked the least. While all predictors showed the same order of magnitude on the biasing the decision to a ``negative'' outcome, we found that the impact of the region was consistently in the first rank.

While this study is limited in its scope and the sample size too small to obtain accurate predictive logistic regression classifiers for other datasets, we believe that the result is an interesting first step. By itself, it already offers insights in the characteristics of the 3-D Secure decision making in the back-end, normally shrouded from the user.

\subsection{Future Work}
So far, we have considered each line of outcomes separately.
Of course, these analyses do not take into account the interplay between dependent variables.
As future work, we anticipate it to be fruitful to analyze 3-D Secure either with 
a multinomial logistic regression or hidden model estimation.

To evaluate the impact of \emph{personalized} user profiles on payment card transactions governed by 3-D Secure, future work could include a large-scale experiment with many participants and a diversity of card payment histories.

\begin{DocumentVersionSTAST}
 \begin{acks}
This work was in parts supported by a grant of the \grantsponsor{NCSC}{National Cyber Security Centre (NCSC)}{https://www.ncsc.gov.uk} through the UK Research Institute in the Science of Cyber Security (RISCS) in its approach to research integrity and reproducibility. Pre-registration, dataset and meta-data annotation are independently hosted at the Open Science Framework (OSF)\footnote{DOI 10.17605/OSF.IO/X6YFH; \url{https://osf.io/x6yfh/}} and publicly available under Creative Commons license. Thomas Gro{\ss} was supported by the \grantsponsor{ERC}{European Research Council (ERC)}{https://erc.europa.eu} Starting Grant ``Confidentiality-Preserving Security Assurance (CASCAde),'' GA n\textsuperscript{o}\grantnum{ERC}{716980}.
\end{acks}
\end{DocumentVersionSTAST}

\begin{DocumentVersionTR}
\section*{Acknowledgements}
This work was in parts supported by a grant of the National Cyber Security Centre (NCSC) through the UK Research Institute in the Science of Cyber Security (RISCS) in its approach to research integrity and reproducibility. Pre-registration, dataset and meta-data annotation are independently hosted at the Open Science Framework (OSF)\footnote{DOI 10.17605/OSF.IO/X6YFH; \url{https://osf.io/x6yfh/}} and publicly available under Creative Commons license. Thomas Gro{\ss} was supported by the ERC Starting Grant ``Confidentiality-Preserving Security Assurance (CASCAde),'' GA n\textsuperscript{o} 716980.
\end{DocumentVersionTR}

\balance

\begin{DocumentVersionSTAST}
\bibliographystyle{ACM-Reference-Format}
\end{DocumentVersionSTAST}
\begin{DocumentVersionTR}
\bibliographystyle{IEEEtran}
\end{DocumentVersionTR}

\bibliography{3DS,methods_resources}


\begin{appendix}
\begin{DocumentVersionSTAST}
\section{Model Evaluation}
\label{sec:model_eval}
\subsection{Logistic Regression: User Challenged}
\label{sec:model_eval_challenged}

\paragraph{Model Selection}
The model with all five predictors was statistically significant, $\chi^2(5) = 21.593$, $p < .001$. We checked for second-level and third-level interactions wrt. the manipulated independent variables and found none.

This model yields an Akaike Information Criterion corrected for small sample sizes of $\const{AIC_c} = 69.73$.
Compared to the best model only including the predictors with the greatest residual drop (\textsf{machine\_data}, \textsf{value}, and \textsf{region}),
this fitted model experiences a small enough information loss ($\Delta = 1.14$) to be classified as having substantial support.

\paragraph{Goodness of Fit}
We computed a likelihood-ratio test to compare the full model selected against the one with minimal \const{AIC_c} (and fewer predictors).
We failed to reject the null hypothesis that the reduced model is true, and, hence, keep the full model. We report the goodness-of-fit in different variants of Pseudo-$R^2$ in Table~\ref{tab:regTableChallengedRefined}. McFadden's $R^2 = .28$.

\paragraph{Diagnostics}
There were two cases with large residuals, but DFbetas were well below $.5$. There were no cases with large leverage.
Assessing for multicollinearity, we found the Variance Inflation Factors (VIFs) all close to $1$, with a mean VIF of $1.09$.

\subsection{Logistic Regression: Transaction Declined}
\label{sec:model_eval_declined}

\paragraph{Model Selection}
The model including all predictors was statistically significant, $\chi^2(5) = 44.409$, $p < .001$. Second-level and third-level interactions between manipulated variables were not statistically significant.

This model comes with a corrected Akaike Information Criterion $\const{AIC_c} = 57.72$.
Compared to the min-$\const{AIC_c}$ model with the predictors \textsf{machine\_data}, \textsf{value}, \textsf{region}, and \textsf{card},
the fitted model has an information loss of $\Delta = 2.3$. 
This is which past the threshold of substantial support, but still considered good evidence.

\paragraph{Goodness of Fit}
Comparing the min-$\const{AIC_c}$ model with the chosen model on goodness-of-fit, we find that a likelihood-ratio test does not reject the null hypothesis.
We report the goodness-of-fit in Pseudo-$R^2$ in Table~\ref{tab:regTableDeclinedRefined}. McFadden's $R^2 = .50$.

\paragraph{Diagnostics}
There were four cases with large residuals, yet DFbetas shown to be below $1$.
There were 12 cases with leverage just touching twice the average leverage, however the DFbetas are consistently less than $1$ and Cook's distance less than $0.1$.
The VIFs are smaller than $2$, where the mean VIF is $1.49$.

\subsection{Logistic Regression: Card Blocked}
\label{sec:model_eval_blocked}

\paragraph{Model Selection}
We have a scenario in which the model including only \textsf{value} and \textsf{region} has the minimal $\const{AIC_c} = 40.39$.

The full model including the three other predictors ($\const{AIC_c} = 44.71$) yields an information loss of $\Delta = 4.32$, having considerably less support. Given the data, this model only carries a likelihood of $1\%$. 

Even though the likelihood-ratio test does not reject the null hypothesis, we consider the model with \textsf{value}, \textsf{region} and\textsf{machine\_data} as robust alternative ($\const{AIC_c} = 41.05$). In comparison with the minimal-$\const{AIC_c}$ model, we have an information loss of $\Delta = 0.67$, yielding substantial evidence. Hence, we select this model.

\paragraph{Goodness of Fit}
We evaluate a likelihood-ratio test to check the goodness-of-fit of the min-$\const{AIC_c}$ model vis-{\`a}-vis the chosen model.
It did not reject the null hypothesis.
Table~\ref{tab:regTableBlockedRefinedInt} contains customary Pseudo-$R^2$ estimates. McFadden's $R^2 = .45$.

\paragraph{Diagnostics}
There was one case with high residuals, but DFbetas smaller than $1$.
There were $9$ cases with a leverage past the double-mean-leverage threshold.
Inspecting DFbetas, we find them to be below $1$, and inspecting the cooks distance, we find it below $0.2$ max.
The VIF was consistently close to 1, with a mean VIF of $1.04$.
\end{DocumentVersionSTAST}

\end{appendix}


\end{document}